\documentclass[twocolumn]{aastex63}

\usepackage[T1]{fontenc}
\usepackage[utf8]{inputenc}
\usepackage{xspace}
\usepackage{grffile}
\usepackage{multirow}
\usepackage{enumitem}
\usepackage{rotating}

\graphicspath{{./}{./figures/}}

\newcommand{\m}[2]{M\,#1#2\xspace}
\renewcommand{\arcsec}{\ensuremath{^{\prime\prime}}\xspace}
\renewcommand{\arcmin}{\ensuremath{^{\prime}}\xspace}
\newcommand{\degree}{\ensuremath{^{\circ}}}
\renewcommand{\deg}{\ensuremath{^\circ}\xspace}
\newcommand{\kms}{km\,s$^{-1}$\xspace}

\newcommand{\jybeam}{Jy\,beam$^{-1}$\xspace}
\newcommand{\mjybeam}{mJy\,beam$^{-1}$\xspace}
\newcommand{\Kkmspc}{K\,km\,s$^{-1}$\,pc$^2$\xspace}

\newcommand{\Msun}{M$_\odot$\xspace}

\newcommand{\Msunpc}[1]{M$_\odot$\,pc$^{-#1}$}

\newcommand{\co}[2]{CO(#1--#2)\xspace}

\newcommand{\gildas}{\textsc{gildas}\xspace}

\newcommand{\alphaCO}{\ensuremath{\alpha_\mathrm{CO}}\xspace}
\newcommand{\alphaCOunit}{M$_\odot$\,(\Kkmspc)$^{-1}$}

\received{-}
\revised{-}
\accepted{-}
\submitjournal{ApJL}

\shorttitle{NOEMA observations of \m82}
\shortauthors{Krieger et al.}

\begin{document}

\title{NOEMA High Fidelity Imaging of the Molecular Gas in and around \m82}

\correspondingauthor{Nico Krieger}
\email{krieger@mpia.de}

\author[0000-0003-1104-2014]{Nico Krieger}
    \affil{Max-Planck-Institut f\"ur Astronomie, K\"onigstuhl 17, 69120 Heidelberg, Germany}
\author[0000-0003-4793-7880]{Fabian Walter}
    \affil{Max-Planck-Institut f\"ur Astronomie, K\"onigstuhl 17, 69120 Heidelberg, Germany}
    \affiliation{National Radio Astronomy Observatory, P.O. Box O, 1003 Lopezville Road, Socorro, NM 87801, USA}
    
\author[0000-0002-5480-5686]{Alberto D. Bolatto}
    \affiliation{Department of Astronomy and Joint Space-Science Institute, University of Maryland, College Park, MD 20742, USA}
\author{Pierre Guillard}
    \affiliation{Institut d’Astrophysique de Paris, 98bis bvd Arago, 75014, Paris, France}
\author[0000-0003-1939-5885]{Matthew Lehnert}
    \affiliation{Sorbonne Université, CNRS UMR 7095, Institut d’Astrophysique de Paris, 98bis bvd Arago, 75014, Paris, France}
\author[0000-0002-2545-1700]{Adam Leroy}
    \affiliation{Department of Astronomy, The Ohio State University, 4055 McPherson Laboratory, 140 West 18th Ave, Columbus, OH 43210, USA}
\author[0000-0003-3061-6546]{Jérôme Pety}
    \affiliation{Institut de Radioastronomie Millimétrique (IRAM), 300 Rue de la Piscine, F-38406 Saint Martin d’Hères, France}
    \affiliation{LERMA, Observatoire de Paris, PSL Research University, CNRS, Sorbonne Universit\'es, 75014 Paris}

\author[0000-0001-6527-6954]{Kimberly L. Emig}
    \affiliation{National Radio Astronomy Observatory, 520 Edgemont Road, Charlottesville, VA 22903-2475, USA}
    \altaffiliation{Jansky Fellow of the National Radio Astronomy Observatory} 
\author[0000-0003-2508-2586]{Rebecca C. Levy}
    \affiliation{Department of Astronomy and Joint Space-Science Institute, University of Maryland, College Park, MD 20742, USA}
\author{Melanie Krips}
    \affiliation{Institut de Radioastronomie Millimétrique (IRAM), 300 Rue de la Piscine, F-38406 Saint Martin d’Hères, France}
\author[0000-0003-4996-9069]{Hans-Walter Rix}
    \affil{Max-Planck-Institut f\"ur Astronomie, K\"onigstuhl 17, 69120 Heidelberg, Germany}
\author[0000-0002-3848-1757]{Dragan Salak}
    \affiliation{Tomonaga Center for the History of the Universe, University of Tsukuba, 1–1–1 Tennodai,Tsukuba, Ibaraki 305–8571, Japan}
\author[0000-0003-4678-3939]{Axel Weiss}
    \affiliation{Max-Planck-Institut f\"ur Radioastronomie, Auf dem H\"ugel 69, 53121 Bonn, Germany}
\author[0000-0002-3158-6820]{Sylvain Veilleux}
    \affiliation{Department of Astronomy and Joint Space-Science Institute, University of Maryland, College Park, MD 20742, USA}


\begin{abstract}

We present a 154 pointing IRAM NOEMA mosaic of the \co10 line emission in and around the nearby starburst galaxy \m82. The observations, complemented by zero--spacing observations, reach a spatial resolution of $\sim$30\,pc ($\sim 1.9\arcsec$) at 5.0\,\kms spectral resolution, sufficient to resolve the molecular gas in the central starburst disk, the outflow, as well as the tidal streamers. The resulting moment and peak brightness maps show a striking amount of structure. Using a clump decomposition algorithm, we analyse the physical properties (e.g., radii $R$, line widths $\sigma$, and masses $M$) of $\sim2000$ molecular clouds. To first order, the clouds' properties are very similar, irrespective of their environment. This also holds for the size--line width relations of the clouds. The distribution of clouds in the $\sigma^2/R$ vs.\ column density $\Sigma$ space suggests that external pressure does not play a significant role in setting their physical parameters in the outflow and the streamers. We find that the clouds in the streamers stay approximately constant in size ($R\,\sim 50$\,pc) and mass ($M\,\sim 10^5$\,\Msun) and do not vary with their projected distance from \m82's center. The clouds in the outflow, on the other hand, appear to decrease in size and mass with distance towards the Southern outflow. The reduction in the molecular gas luminosity could be indicative of cloud evaporation of embedded clouds in the hot outflow.

\end{abstract}

\keywords{galaxies: individual (M82), galaxies: ISM, galaxies: starburst, ISM: clouds}

\section{Introduction}\label{section:introduction}

Galaxy--wide outflows driven by star formation are thought to be crucial drivers in galaxy evolution. Stellar feedback caused by intense central star formation activity can launch such outflows, leading to significant fractions of baryons (ionized, atomic, and molecular gas) that escape the main body of the galaxy \citep[e.g.][]{2020A&ARv..28....2V}. Outflows in starburst galaxies are a multi--phase phenomenon and have been observed across the electro--magnetic spectrum from X-ray \citep[e.g.][]{2007ApJ...658..258S}, UV \citep[e.g.][]{2005ApJ...619L..99H}, optical H$\alpha$ \citep[e.g.][]{1998ApJ...493..129S,2009ApJ...696..192W},  IR \citep[e.g.][]{2009ApJ...700L.149V}, and cold dust \citep{2010A&A...518L..66R}, PAH emission \citep[e.g.][]{2006ApJ...642L.127E}, warm H$_2$ \citep[e.g.][]{2015MNRAS.451.2640B}, to  (sub--)millimeter and radio emission \citep[e.g.][]{2002ApJ...580L..21W,2013Natur.499..450B,2015ApJ...814...83L,2018ApJ...856...61M}. 
While evidence for galactic outflows is manifold, a detailed characterization is restricted to only a few local systems, where the relevant processes can be spatially resolved at high sensitivity.
In particular, the physical  characterization of the outflowing gas mass is important as it influences a galaxy's ability to form stars in the future. In this context the molecular gas phase is particularly relevant because it often carries the dominant mass fraction of all baryons \citep[e.g.][]{2019ApJ...881...43K}. The fate of the molecular gas in the outflow is itself not clear. Outflowing molecular clumps may be shocked and evaporated by the fast and hot outflowing gas becoming part of the hot phase \citep{2015ApJ...805..158S,2017ApJ...834..144S}, or they may act as condensation seeds that gain mass, momentum, and velocity by strongly cooling hot gas \citep{2018MNRAS.480L.111G,2020ApJ...894L..24F,Abruzzo:2021uv}. Verifying which of these possibilities actually takes place in molecular outflows would strongly impact the interpretation of the observational data, as well as our understanding of the physical processes driving the cool phases of galaxy outflows.

\m82 is one of the few galaxies that show an extended dusty outflow \citep[e.g.][]{2006ApJ...642L.127E,2009ApJ...700L.149V,2013PASJ...65...66S,2015MNRAS.451.2640B,2016ApJ...830...72C}. Because of its close proximity ($\mathrm{D} = 3.5$\,Mpc) and its almost edge--on orientation \citep[inclination $i \sim 80\deg$; e.g.][]{1993A&A...272...98M}, \m82 offers a unique laboratory to study galactic winds. CO emission associated with the outflow has first been indicated by observations at the Nobeyama 45\,m telescope by \citet{1987PASJ...39..685N} and has been confirmed in \citet{2001ApJ...562L..43T} based on observations with the Five College Radio Astronomical Observatory (FCRAO) single dish telescope.
A more detailed view of the central disk of \m82 was achieved by the first interferometric map of \m82's CO emission \citep[using the millimeter interferometer of the Owen's Valley Radio Observatory, OVRO;][]{2002ApJ...580L..21W}.
More recently, a wide--field single--dish CO map (obtained at the IRAM 30\,m telescope) demonstrated that the molecular outflow is indeed as extended as observed in other tracers (H$\alpha$, X--rays), out to distances of 3\,kpc \citep{2015ApJ...814...83L}. 
These previous single--dish studies could constrain the overall dynamics and the amount of outflowing molecular gas, but given their effective resolution of $\gtrsim 300$\,pc, they could not resolve the actual structure within the outflow.

To characterize the spatial structure of the molecular outflow in \m82, we obtained high-resolution observations over a large field-of-view in \m82 with the Northern Extended Millimeter Array (NOEMA). These interferometric observations were complemented with zero spacing information from the IRAM 30\,m telescope. 
Compared to previous interferometric observations \citep{2002ApJ...580L..21W}, we achieve a sensitivity that is 3 times deeper, a synthesized beam area that is 7 times smaller, and cover an area on the sky that is 3 times larger. At a distance to \m82 of 3.5\,Mpc, one arcsec corresponds to 17.0\,pc (i.e. one arcmin to 1.02\,kpc)

\begin{figure*}
    \centering
    \includegraphics[width=\linewidth]{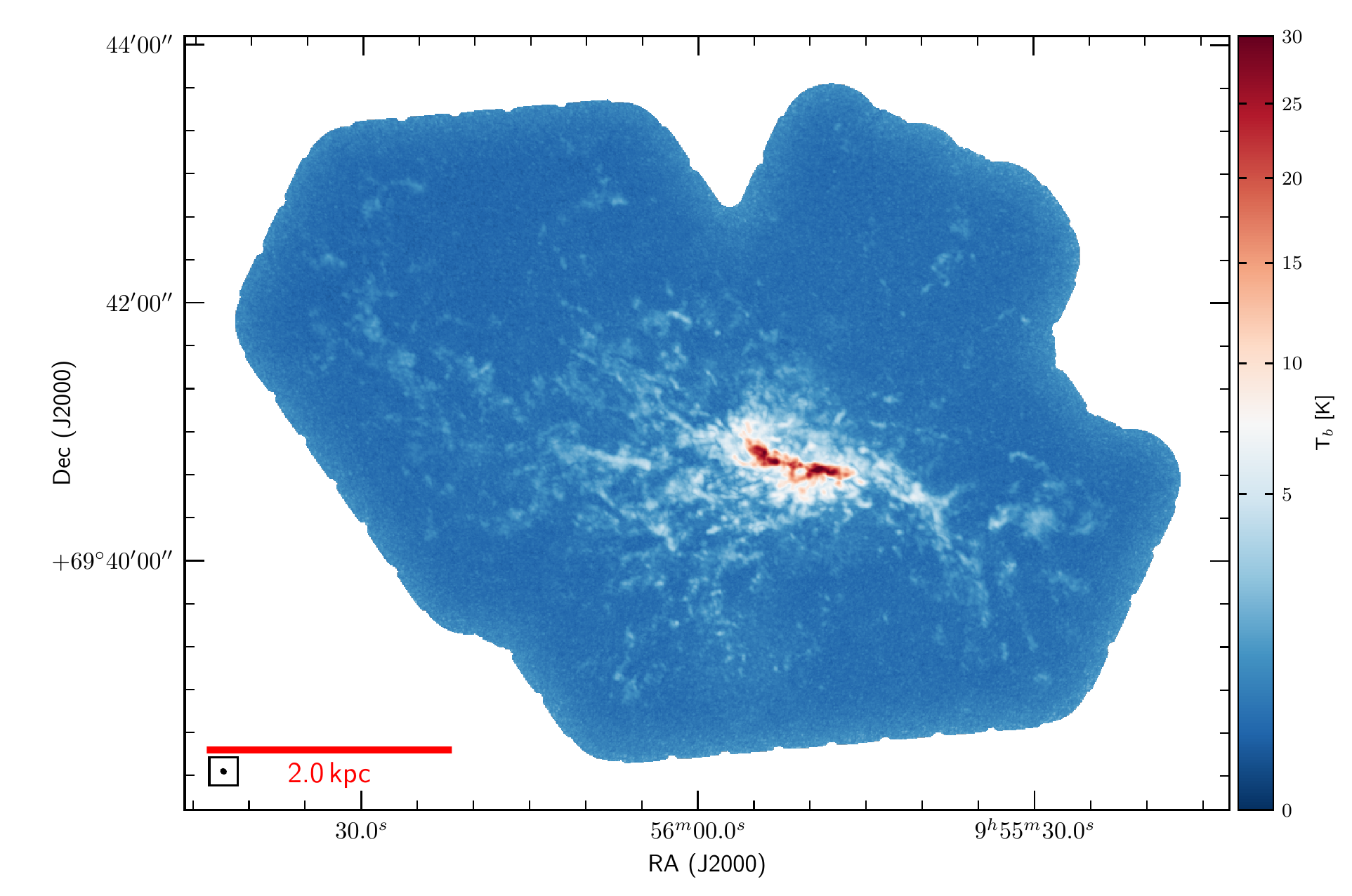}
    \caption{NOEMA \co10 mosaic of \m82. The colormap, showing the peak main beam intensity, is chosen to highlight the visibility of the faint emission in the molecular streamers and outflows, which leads to saturation of the brightest emission in the central disk (at $\sim 35$\,K). The synthesized beam ($\sim\,30$\,pc) is shown in the bottom left corner.
    \label{figure: CO peak}}
\end{figure*}

In this paper, we describe the observations and data reduction in Sec.~\ref{section: observations}, and present the imaging of the \co10 data in Sec.~\ref{section: data}. We then characterize the small scale structure of the molecular gas in the outflow, \m82's disk, as well as the surrounding tidal features in Sec.~\ref{section: discussion}. Finally, our results are summarized in  Sec.~\ref{section: summary}.

\section{Observations}\label{section: observations}

\subsection{NOEMA}\label{section: NOEMA}

The NOEMA mosaic covers an area of $\sim 25$\,arcmin$^2$ over 7.7\arcmin (7.9\,kpc) along the major axis and out to $\pm2.8$\arcmin ($\pm2.9$\,kpc) along the minor axis of \m82. This area covers the regions in the disk and the outflows/streamers that host significant CO emission ($I_\mathrm{CO(2-1)} > 1.5$\,K\,\kms) as mapped by single--dish observations \citep{2015ApJ...814...83L}. The mosaic consists of 154 pointings with a hexagonal arrangement with half-width primary beam spacing (21.5\arcsec), to achieve an approximately uniform sensitivity across the field–of–view. While the main focus of the observations is the rotational ground state transition of carbon monoxide (\co10 at $\nu_\mathrm{rest} = 115.271$\,GHz), the wide bandwidth correlator at NOEMA also covers other molecular lines such as $^{13}$\co10, CS(2-1) or CN(1-0) that are not discussed in this paper.

The NOEMA observations were carried out under project ID W18BY between March 2019 and January 2020. Out of the 38 observed runs, 4 had to be dropped entirely due to poor atmospheric conditions yielding 34 runs with a combined 44~hours of on-source time using an equivalent 10-antenna array.
70\% of the total time was observed in NOEMA's C configuration, and 30\% in its D configuration.
LKHA101 or MWC349 were used as  flux calibrators and the complex gain calibration was performed on 0836+710 and 0954+658.
We tuned the correlator PolyFix to cover the frequency ranges $92.6-100.3$\,GHz and $107.8-115.6$\,GHz at $2.03$\,MHz spectral resolution.

\subsection[IRAM 30m]{IRAM 30\,m}\label{section: 30m}

Short-spacing data were observed with the IRAM 30\,m telescope in two runs in April (28/29/30 Apr) and June/July (30 Jun, 01/02/03 Jul) of 2020 for a total of 22~hours of on-source time.
The focus and pointing calibrators were 0954+658, 0923+392 and 0836+710.
The area of the NOEMA mosaic was covered by a $560\arcsec \times 560\arcsec$ on-the-fly map using two sets of scans along right ascension and declination. 
The Eight MIxer Receiver (EMIR) was tuned to cover the frequency ranges $94.4-102.4$\,GHz and $110.5-118.1$\,GHz which includes the \co10 line.

\subsection{Data reduction and imaging}\label{section: imaging}

Calibration of the NOEMA interferometric data was done in \gildas (Version \texttt{jul20a}) using the \textsc{clic} tool.
The data were calibrated using the standard pipeline. 
From the calibrated visibilities, a subset of 800\,\kms width at 5.0\,\kms resolution around the \co10 line was extracted for further analysis, as presented in this paper.
Continuum emission was fitted in line--free channels away ($>|\pm 250\,$\kms$|$) from the line center at 210\,\kms systemic velocity. The obtained continuum fit was then subtracted from each visibility to provide a line--only dataset. Self--calibration slightly improved the image fidelity in three of the central pointings that contain bright emission, i.e., most of \m82's disk. In these cases, we applied the self--calibration solutions after four iterations.

The 30\,m single dish observations were also reduced in \gildas (Version \texttt{jul20a}) using the \textsc{mrtcal} and \textsc{class} tools.The \textsc{mrtcal} tool automatically removed the atmospheric contribution and calibrated the intensity scale in unit of antenna temperature. In \textsc{class}, we first extracted a frequency window of 500\,MHz centered around the rest frequency of  115271.203\,MHz from each spectrum and we converted the intensity scale to main beam temperature using using the task \texttt{beam\_efficiency /ruze}. We then subtracted a first order spectral baseline fitted on line-free channels from the line center (same velocity ranges as for the interferometer data) and filtered out all spectra whose baseline noises are larger than three times the standard deviation of the noise distribution. We then resampled the spectra to the same spectral grid as the NOEMA data. Finally, the spectra were gridded through convolution with a Gaussian kernel whose FWHM is one third the natural resolution of the IRAM-30\,m telesecope using the \texttt{xy\_map} task. We choose as projection center the phase center of the NOEMA mosaic and a pixel size of $4\times4''$. Visual inspection of the obtained position-position-velocity cube reveals well-behaved intensity and noise distribution.

We imaged the NOEMA data first and then we combined the single-dish and interferometer maps with the \textsc{casa} \texttt{feather} tool.
We used the \textsc{mapping} tool in the parallelized version of \gildas (Version \texttt{feb20a}) to speed up imaging and deconvolution by a factor proportional to the number of cores. In an effort to retain as much of the faint emission away from the central disk as possible, we produced interferometer-only images with natural weighting. As the synthesized beam slightly varies over the field of view, we regularized the deconvolved flux with a Gaussian clean beam whose FWHM was choosen as the largest measured synthesized beam. This gives a uniform spatial resolution of $2.08\arcsec \times 1.65\arcsec$ at a position angle of $51.2\deg$ ($0.4\arcsec \times 0.4\arcsec$ pixels). This corresponds to 36.3\,pc $\times$ 28.9\,pc at the distance of \m82.
We cleaned the emission (using the Steer clean algorithm, \texttt{sdi}) to an absolute flux level (\texttt{ares}) of 3.0\,\mjybeam ($\sim 0.6\sigma$) with as many iterations as required (\texttt{niter\,=\,0}) inside a clean mask.
Corrections for the primary beam response pattern are applied.
For feathering, we converted the single dish map to flux density units and run \texttt{feather} in \textsc{casa} (Version 5.6.1-8).
The conversion from flux density to brightness temperature in the final combined map was done using a conversion factor of 26.8\,K\,(\jybeam)$^{-1}$.

The median root-mean-square (RMS) noise per pixel in a 5.0\,\kms--wide channel of the final cube is 138\,mK (5.15\,\mjybeam).
We achieve a typical noise level of $100-150$\,mK over the entire map, except for a single pointing towards the south that lacks observing time causing locally increased noise values of $\sim 250$\,mK. For all further analysis, we mask out the edge of the mosaic at an offset of $\sim 16\arcsec$ from the map edge.

\section{Data presentation}\label{section: data}

\subsection{Data cube}\label{section: channel maps}

\begin{figure*}
    \centering
    \includegraphics[width=0.49\linewidth]{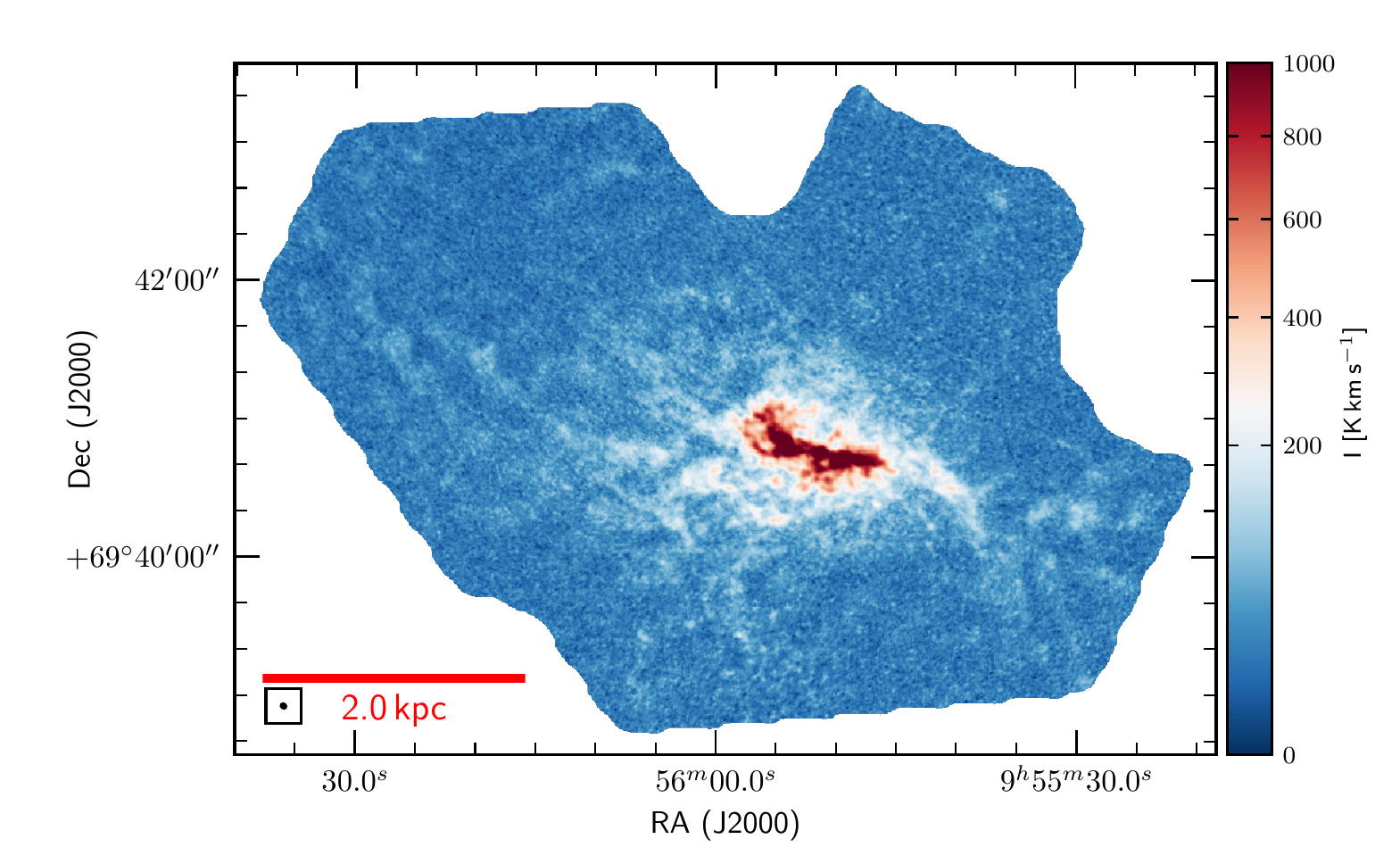}\hfill
    \includegraphics[width=0.49\linewidth]{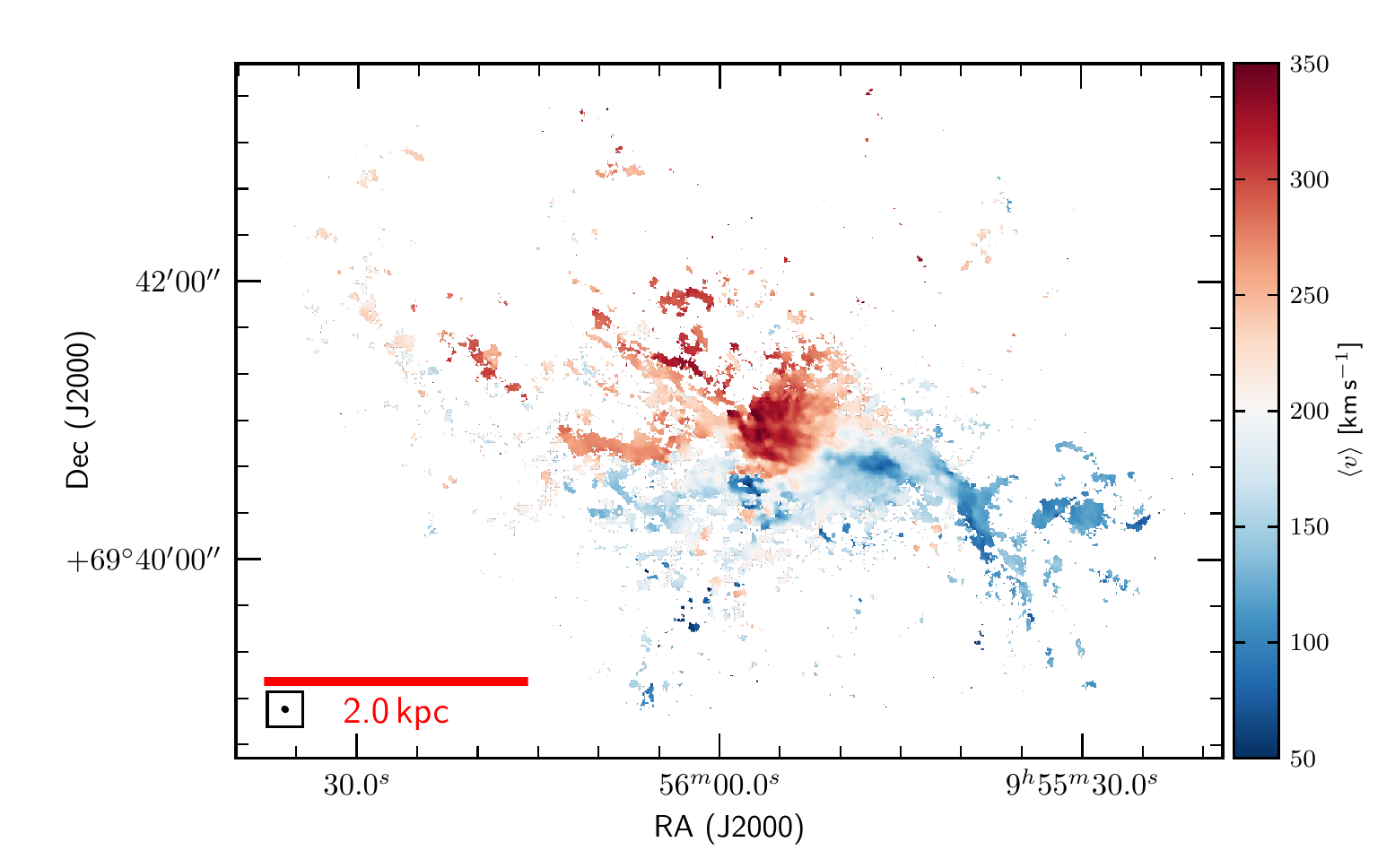}
    \\ \vspace*{2mm}
    \includegraphics[width=0.49\linewidth]{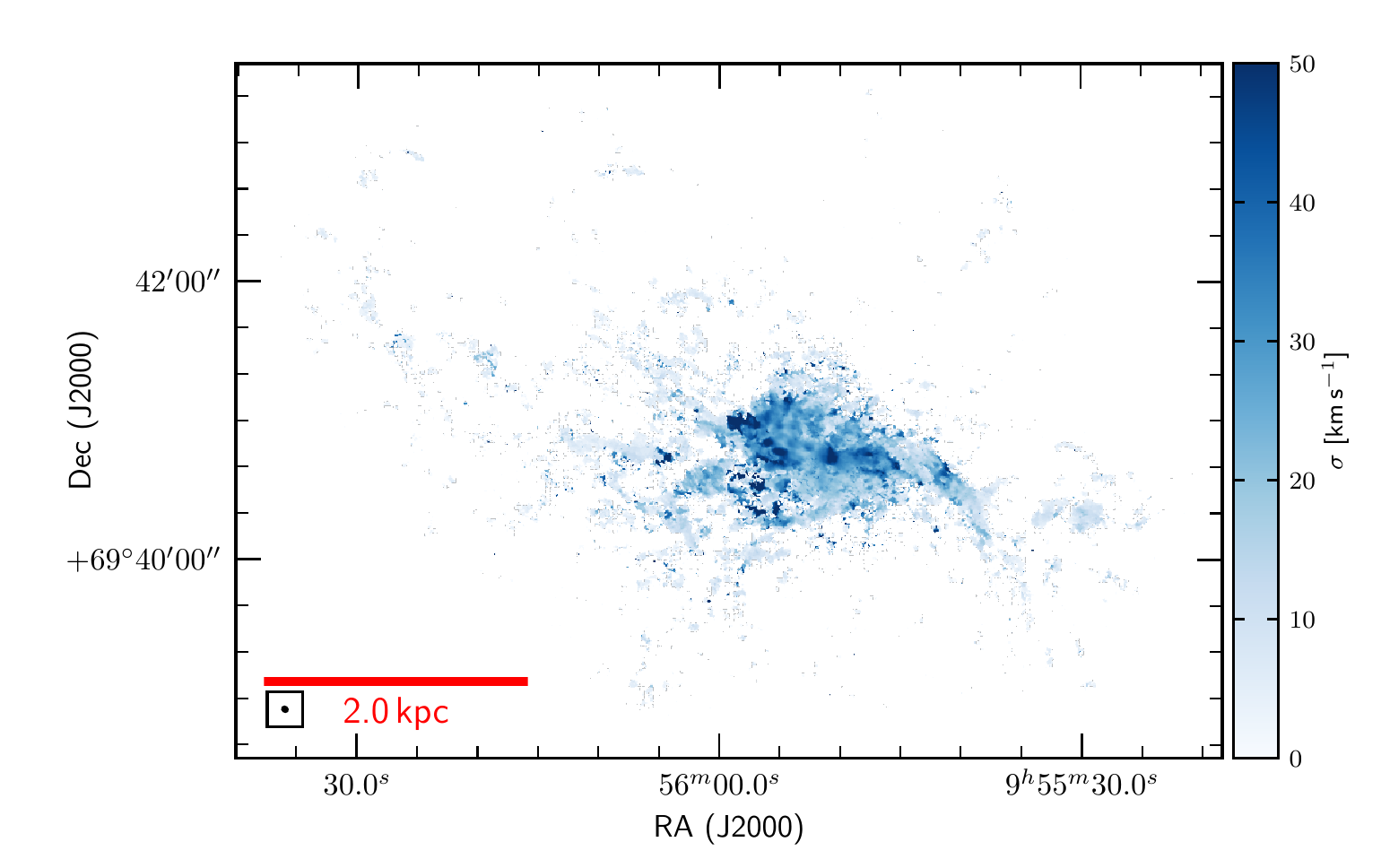}\hfill
    \includegraphics[width=0.49\linewidth]{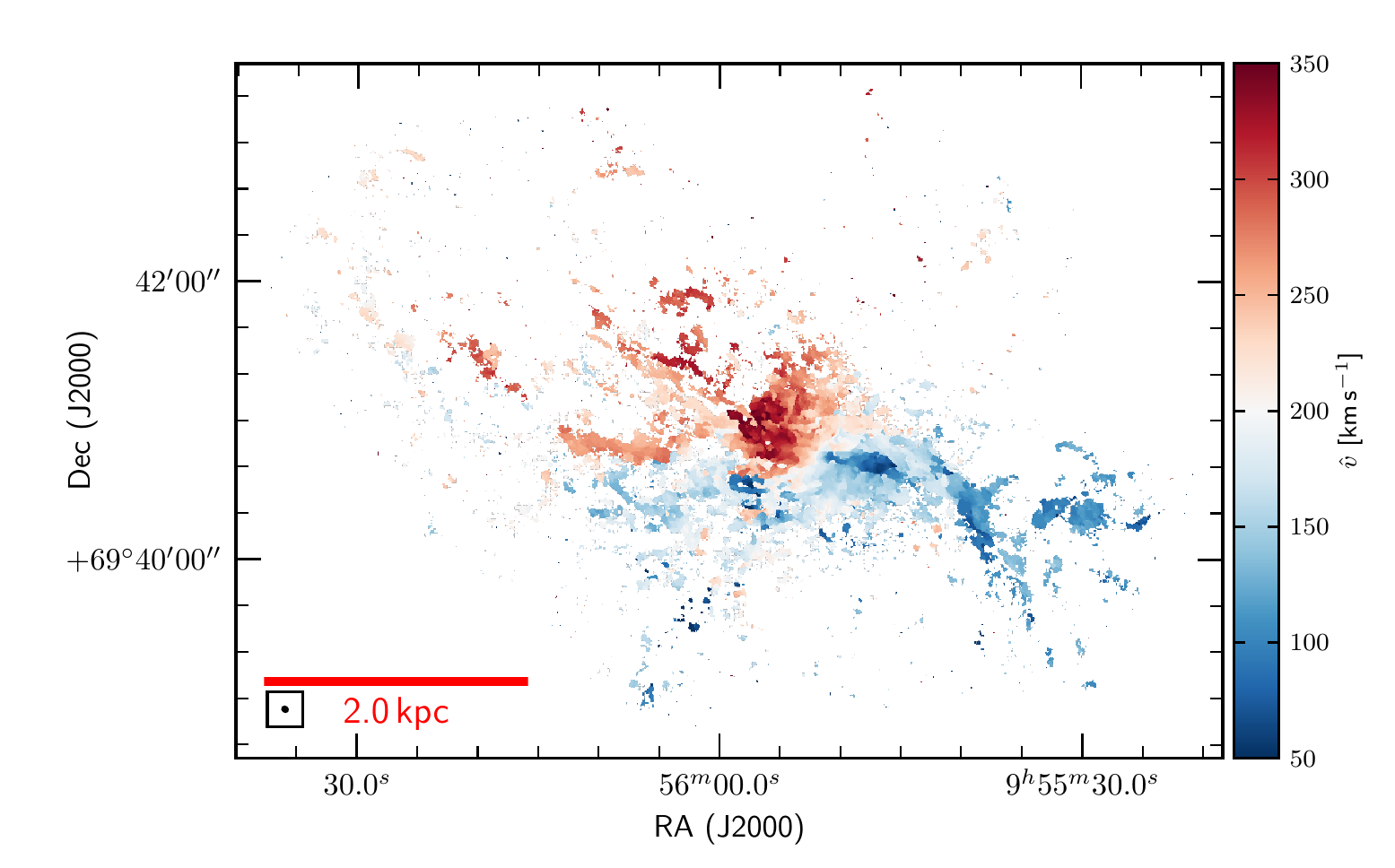}
    \caption{\co10 moment maps of \m82: integrated intensity map (moment 0, top left), intensity-weighted velocity map  (moment 1, top right), and intensity-weighted velocity dispersion map (moment 2, bottom left) of \m82. We also show the velocity at peak \co10 intensity in the bottom right panel. The colormap of the moment 0 map is chosen to increase the visibility of the fainter emission in the molecular streamers and outflow, leading to saturation of pixels in the central disk (peak $\sim 3000$\,K\,\kms). All kinematic maps were created from the \co10 data cube blanked at a $\mathrm{SNR} = 5$ threshold. The systemic velocity is $\sim 210$\,\kms. The synthesized beam ($\sim$\,30\,pc, $\sim 1.9\arcsec$) is shown in the bottom left corner of each map.
    \label{figure: CO maps}}
\end{figure*}

\begin{figure*}
    \centering
    \includegraphics[width=\linewidth]{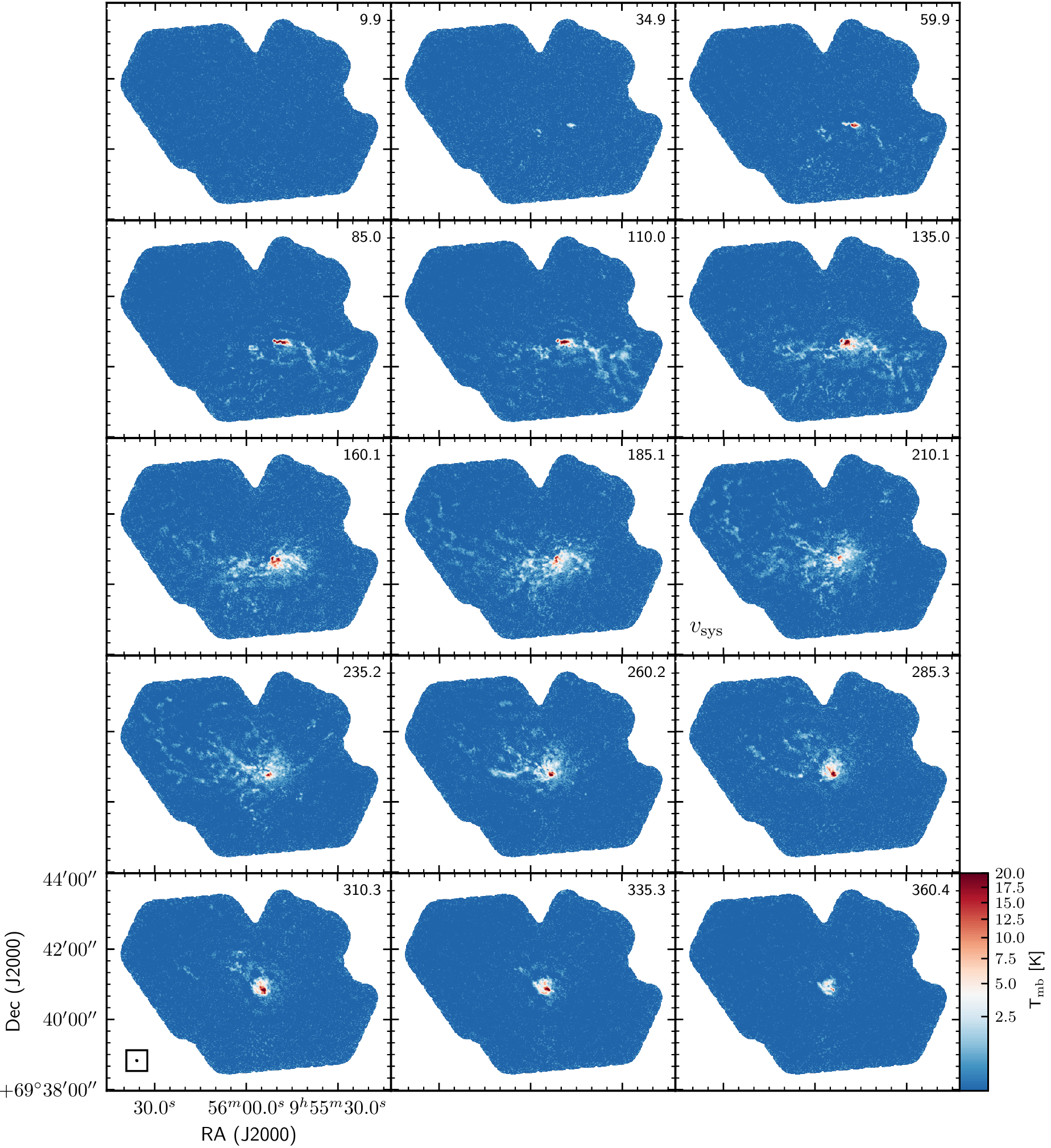}
    \caption{Channel maps of the \co10 emission in \m82, where we show only every $5^\mathrm{th}$ channel of 5.0\,\kms width (the corresponding velocity in \kms in the top right corner of each panel). As for the peak intensity map (Fig.~\ref{figure: CO peak}), the colormap is chosen to increase the visibility of the fainter emission in the molecular streamers and outflows, which leads to saturation of the brightest emission in the center ($\sim 35$\,K). The synthesized beam ($\sim$\,30\,pc) is shown in the bottom left panel. The increase of noise towards the center of the most southern pointings is due to one pointing that has decreased sensitivity.
    \label{figure: CO channel map}}
\end{figure*}

\begin{figure*}
    \centering
    \includegraphics[width=\linewidth]{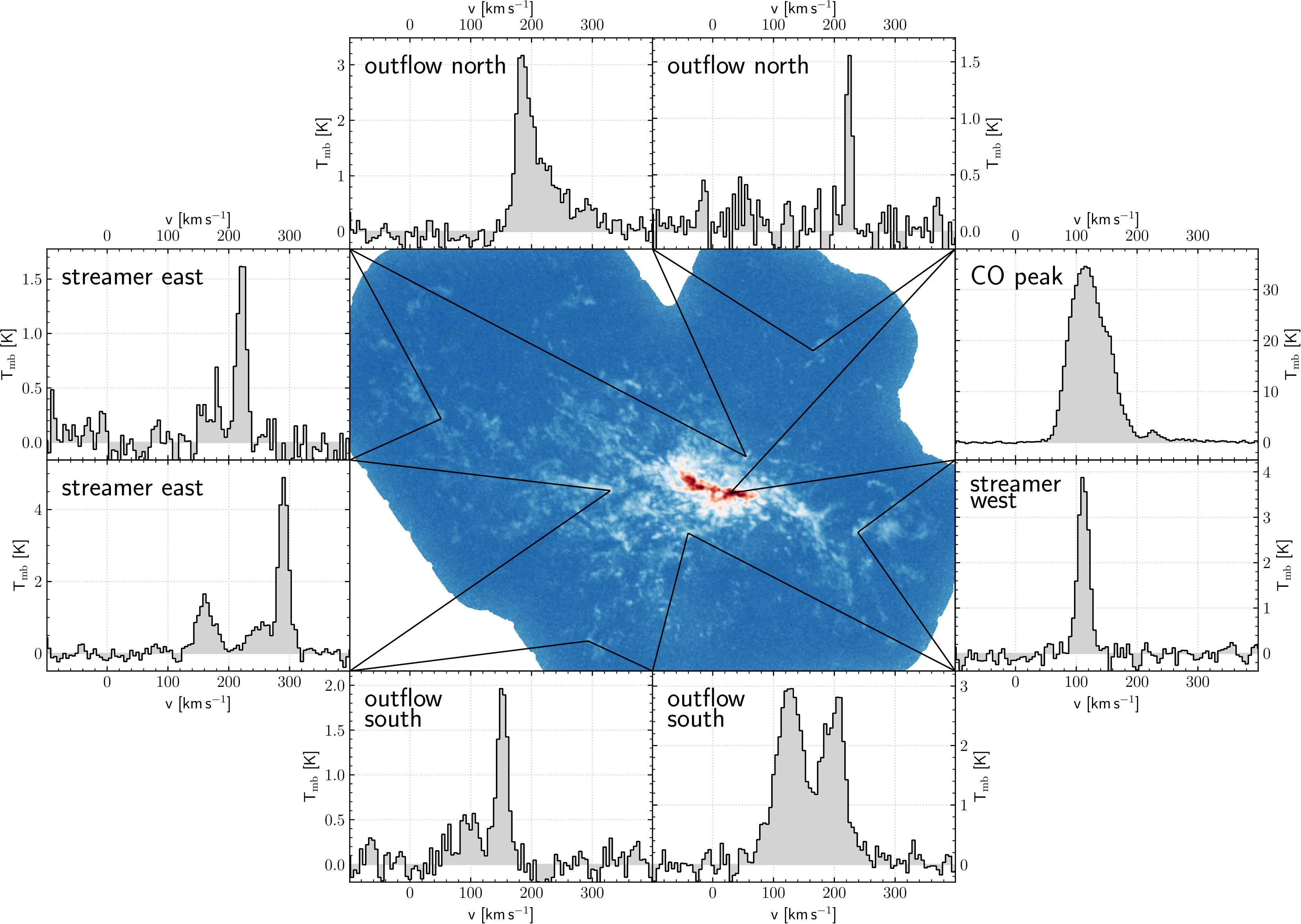}
    \caption{Single pixel CO spectra of selected position. The background CO peak intensity map (Fig.~\ref{figure: CO peak}) is shown to indicate the locations of the spectra. Note that the scale on the y--axis of the spectra changes as a function of position.
    \label{figure: CO spectra}}
\end{figure*}

In Fig.~\ref{figure: CO peak}, we present the \co10 peak brightness map of the NOEMA \m82 mosaic. Fig.~\ref{figure: CO maps} shows the integrated intensity (moment 0), intensity-weighted velocity (moment 1), intensity-weighted velocity dispersion (moment 2) and velocity at peak intensity. At our spatial resolution of $\sim$\,30\,pc, the observations reveal a high degree of filamentary and clumpy structure of the molecular gas emission in and around the central starburst disk (see discussion in Sec.~\ref{section: discussion}). The maps also show a complex velocity structure beyond the central rotating disk.

The \co10 channel maps are shown in Fig.~\ref{figure: CO channel map}. As is already evident from Fig.~\ref{figure: CO peak}, \co10 is detected across the full area mapped by our observations. This includes the central starburst disk,  the northern and southern outflow cones, as well as two tidal arms (`streamers') towards the east and west (see Sec.~\ref{section: region definition} and Fig.~\ref{figure: region definition}).
For reference, some CO spectra are shown in Fig.~\ref{figure: CO spectra}.

Both the peak moment map (Fig.~\ref{figure: CO peak}), as well as the channel maps (Fig.~\ref{figure: CO channel map}) confirm the velocity asymmetry of the molecular outflow, however now on smaller spatial scales than seen before \citep{2002ApJ...580L..21W,2015ApJ...814...83L}. The northern and southern outflows are not symmetric with respect to the galaxy's center \citep[see also e.g.,][]{1998ApJ...493..129S}: while the northern outflow shows signs of rotation following that of the central disk, the southern outflow appears to be blue--shifted and does not feature a clear velocity gradient parallel to the major axis of the disk. This may be caused by contamination of the the northern outflow by the foreground disk. The southern outflow breaks up into individual clouds at approximately constant velocity $v \sim 150$\,\kms, interspersed with some clouds at significantly higher ($v \sim 250$\,\kms, red) and lower ($v \sim 75$\,\kms, dark blue) velocities. This large range of velocities is significantly higher than the dispersion within an individual cloud ($<\,10\,$\kms), and suggests that these molecular clouds are associated with the front and back of the ionized/neutral outflow cone (suggested by observations of the ionized outflow, e.g., \citealt{2020ApJ...904..152L}). We see a similar range of velocity offsets, albeit less pronounced, in the molecular clouds associated with the northern outflow.

The tidal streamers towards the east and the west connect smoothly to the outer disk, but show complex internal structure. The bulk of the gas does not display a strong velocity gradient along the streamers out to distances of several kpc from the center. However, as in the outflow regions, there are some regions where the velocities of individual neighboring clouds (in projection)  differ by up to $\sim 150$\,\kms in the eastern streamer. We attribute these projected velocity differences to distinct velocity components created by the tidal interaction. Towards the south-east, the velocity structure is complex as the eastern streamer ($v \sim 200$\,\kms) and the southern outflow ($v \sim 150$\,\kms) start to overlap in projection.

\subsection{Region definition}\label{section: region definition}

\begin{figure}
    \centering
    \includegraphics[width=\linewidth]{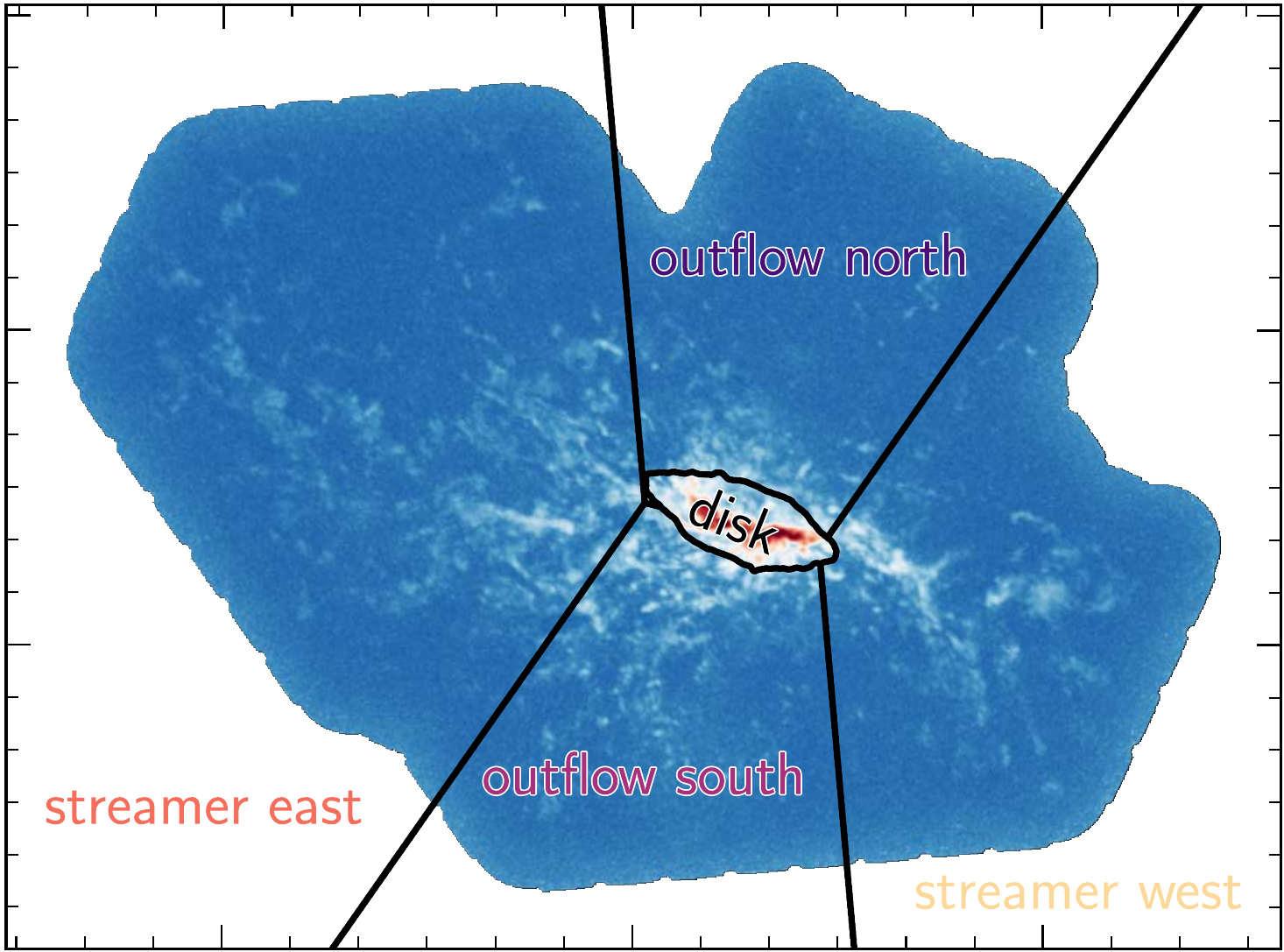}
    \includegraphics[width=\linewidth]{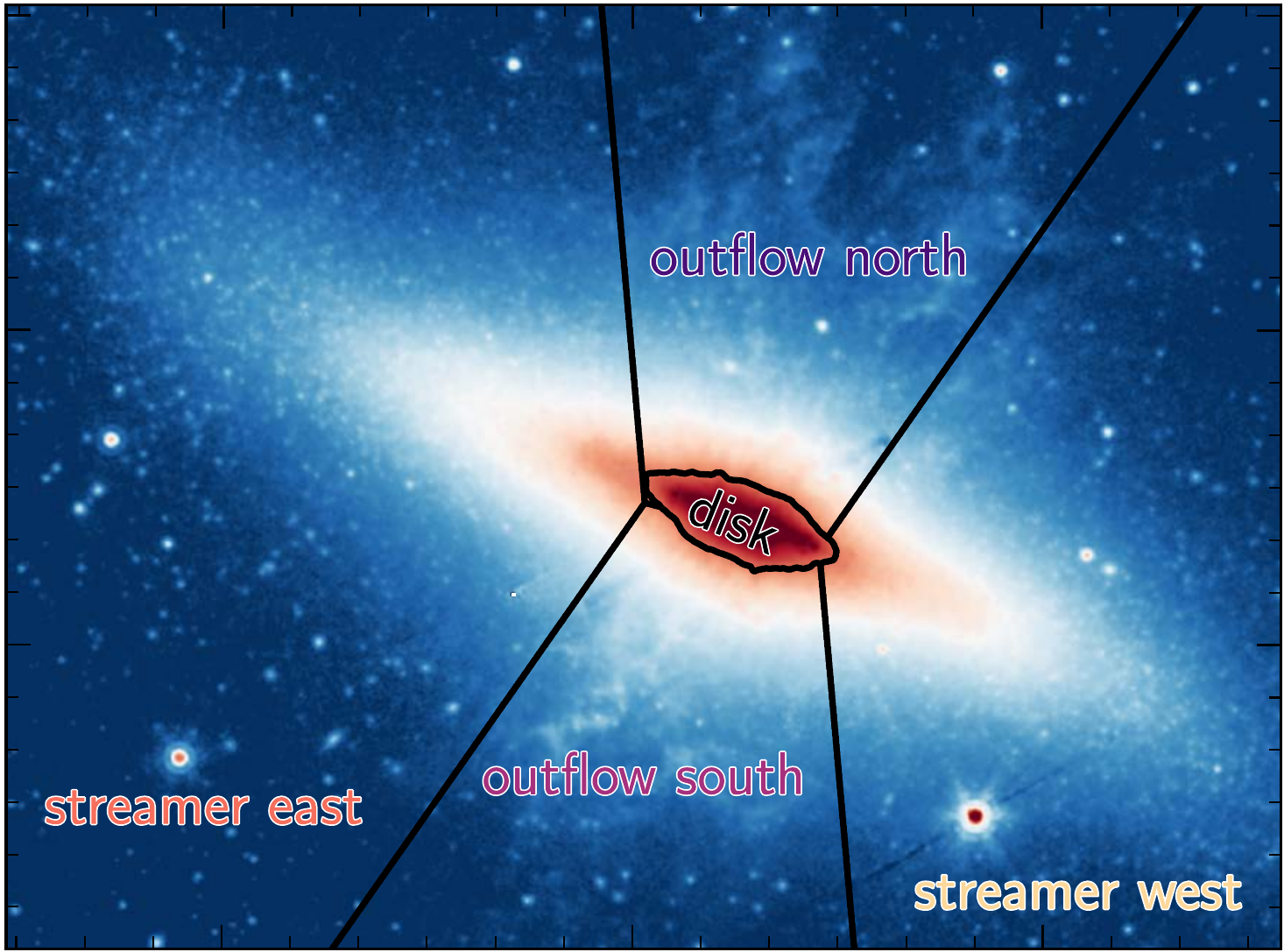}
    \caption{Region masks for the analysis in Sec.~\ref{section: discussion} compared to the NOEMA \co10 map (\emph{top}) and the IRAC band~2 image $4.5\,\mu$m (\emph{bottom}). We separate \m82 into five regions, focusing on the disk, outflows (north and south) and tidal streamers (east and west), as indicated in the figures. North is up, east is to the left (cf. Fig.~\ref{figure: CO peak}.)
    \label{figure: region definition}}
\end{figure}

For further discussion, we here define five regions that trace the different environments in and around \m82: the disk, the outflows as well as tidal streamers (broadly following \citealt{2002ApJ...580L..21W}). This is shown in Fig.~\ref{figure: region definition} where we overplot these regions on top of the \co10 peak intensity (molecular gas) and IRAC $4.5\,\mu$m (tracing old stars as well as hot dust and NIR line emission in the outflow) maps.

\emph{Disk:} We define \m82's disk region based on its stellar disk, adopting a cut that is brighter than 20\,MJy/sr in the IRAC band 2 observations \citep[SINGS,][]{2003PASP..115..928K}. 

\emph{Outflow:} We define the outflow regions as two flat symmetric cones (biconical frustum) with a 40\degree opening angle and a width of 600\,pc at the base (excluding the area assigned to the disk), following previous studies of ionized and molecular tracers that have suggested opening angles in the range $\sim30\degree$ to $\sim60\degree$ for the outflow \citep{1988Natur.334...43B,1995A&A...293..703M,1990ApJS...74..833H,1998ApJ...493..129S,2001ApJ...552..133S,2002ApJ...580L..21W,2006ApJ...642L.127E,2009AJ....137..517L,2010A&A...514A..14K,2015ApJ...814...83L}.

\emph{Streamers:} The remaining areas towards the east and west of the outflow and disk consist of the disturbed outer disk and two tidal streamers. Following \citet{2002ApJ...580L..21W}, we denote these as the eastern and western tidal streamers, respectively. These structures are believed to be tidal in nature: the western streamer connects to the neighboring galaxy M\,81, whereas the eastern streamer points in the opposite direction. These streamers are also seen in \ion{H}{1} imaging of significantly larger fields \citep[e.g.,][]{2018ApJ...865...26D}.

\subsection{Second Moment Map}\label{section: pixel statistics}

We show the second moment map of the \co10 emission in Fig.~\ref{figure: CO maps} (bottom left). This map was created after blanking individual channels at a $\mathrm{SNR} \leq 5$ threshold\footnote{If the spectral lines were Gaussian in nature, the second moment would be identical to the velocity dispersion $\sigma$.}.
We caution the reader that our intrinsic velocity resolution is 5.0\,km\,s$^{-1}$. Therefore, line widths smaller than this value will have significant uncertainties, as discussed in the deconvolution analysis below. 

The central starburst disk displays high velocity dispersion values of typically $20-39$\,\kms ($16^{th}$ to $84^{th}$ percentiles) with a median of $\sigma = 29$\,\kms (mean 30\,\kms), but has peaks exceeding $60$\,\kms. The molecular gas that is associated with the northern and southern outflows show significantly lower values, with median velocity dispersions of 11\,\kms (mean 15\,\kms) and 12\,\kms (mean 16\,\kms), respectively. The eastern and western streamers are found at even lower dispersion with medians of 5\,\kms (mean 8\,\kms) and 8\,\kms (mean 11\,\kms), respectively. The transition regions between the disk and the southern outflow, as well as the disk and the eastern streamer show extended areas of enhanced velocity dispersion. An inspection of the channel maps shows that this is caused by overlapping structures with distinct kinematic components. In the following we discuss the properties of the individual molecular clouds in more detail.

\section{Properties of Molecular Clouds}\label{section: discussion}

With $30$\,pc resolution and $5$\,\kms channels, we identify and characterize structures with sizes similar to individual Galactic giant molecular clouds (GMCs) \citep[e.g., see][]{Bolatto:2008iv}. In this section, we decompose our cube into individual compact structures, which we call ``clouds'' for convenience. We measure the size, line width, and luminosity of each cloud and use these to estimate a series of closely related physical properties. Then we measure how the properties of these clouds vary among the different environments in and around \m82 and as a function of projected distance from the central starburst. Because we focus on comparative analysis, many of the uncertainties related to calculating physical quantities effectively ``divide out'' of the analysis. However, we offer a general caution that the absolute values of mass, surface density, and size--line width coefficient implied by our measurements have significant associated systematic uncertainty.

\subsection{Cloud decomposition}\label{section: Fellwalker}

We decomposed the \co10 data cube using `Fellwalker' \citep{Berry:2014ha}, a watershed algorithm, to segment data into a collection of clouds.
For each pixel above a background level, Fellwalker follows the steepest gradient to a local maximum. All pixels that end up at the same maximum define one cloud. It has been demonstrated that, compared to other cloud identification algorithms, Fellwalker shows high completeness and accuracy \citep{2020RAA....20...31L} and we find the results agree well with what we expect by eye. 

We ran Fellwalker using the pyCupid\footnote{\url{https://pycupid.readthedocs.io/en/latest/}} implementation on the NOEMA \m82 \co10 data cube. We start the search at the rms noise level (\texttt{noise}) whereas valid peaks must reach two times the noise (\texttt{minheight}). Each cloud's volume must be greater than two times the beam area times the channel width to be considered a distinct cloud (\texttt{minpix}). The dip between two peaks (\texttt{mindip}) must be larger than the noise level for them to be considered disjoint and the distance must be more than four pixels away from each other (\texttt{maxjump}). These seemingly low thresholds of a few times the noise level allow us to quantify clouds in the fainter regions of the outflow and streamers. In these regions, the signal-to-noise ratio per individual pixel per channel is typically low ($SNR \lesssim 3$) but the ensemble of dozens of such pixels are significant. 

\begin{figure*}
    \centering
    \includegraphics[scale=0.5]{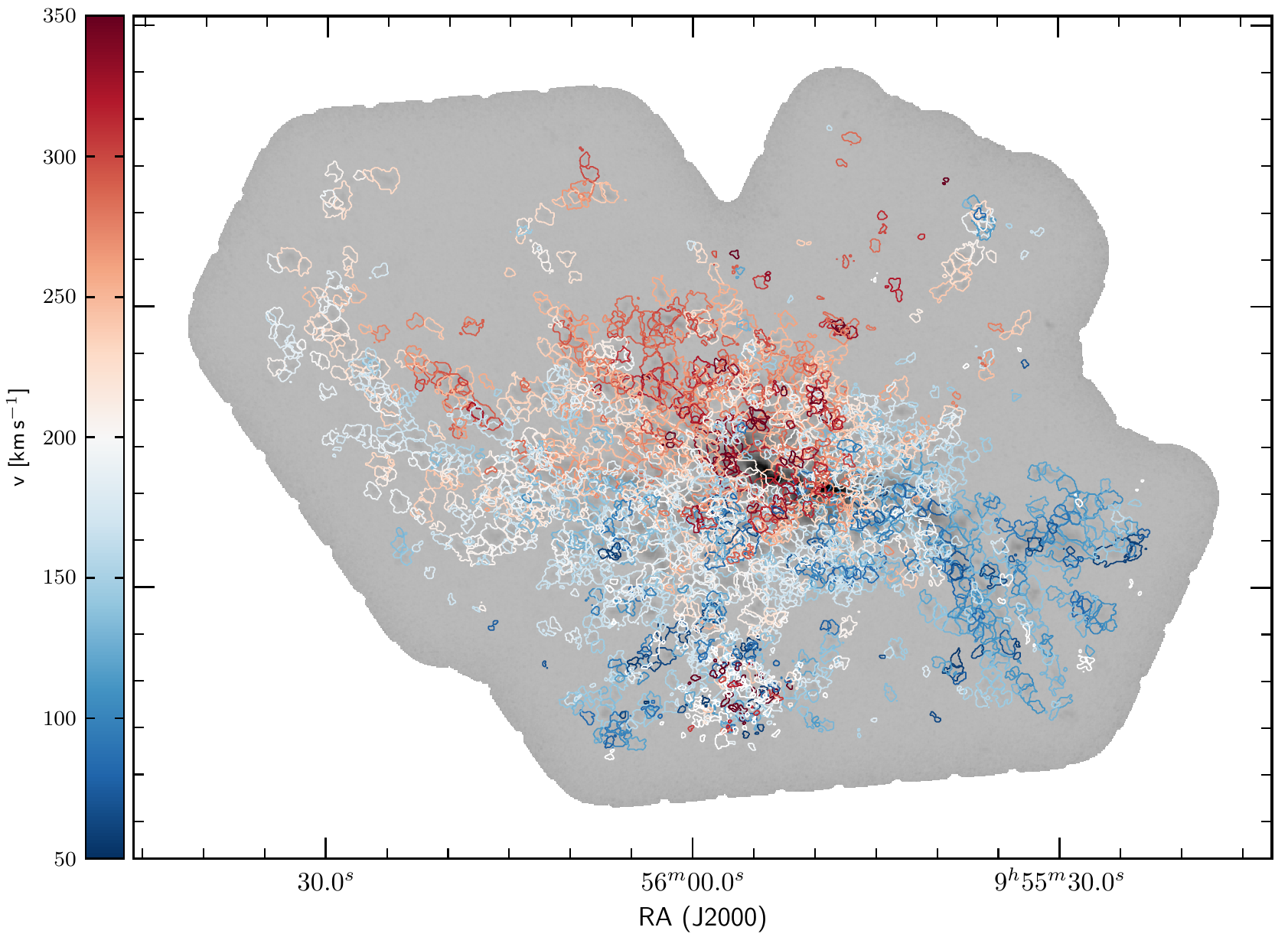}\hfill
    \includegraphics[scale=0.5]{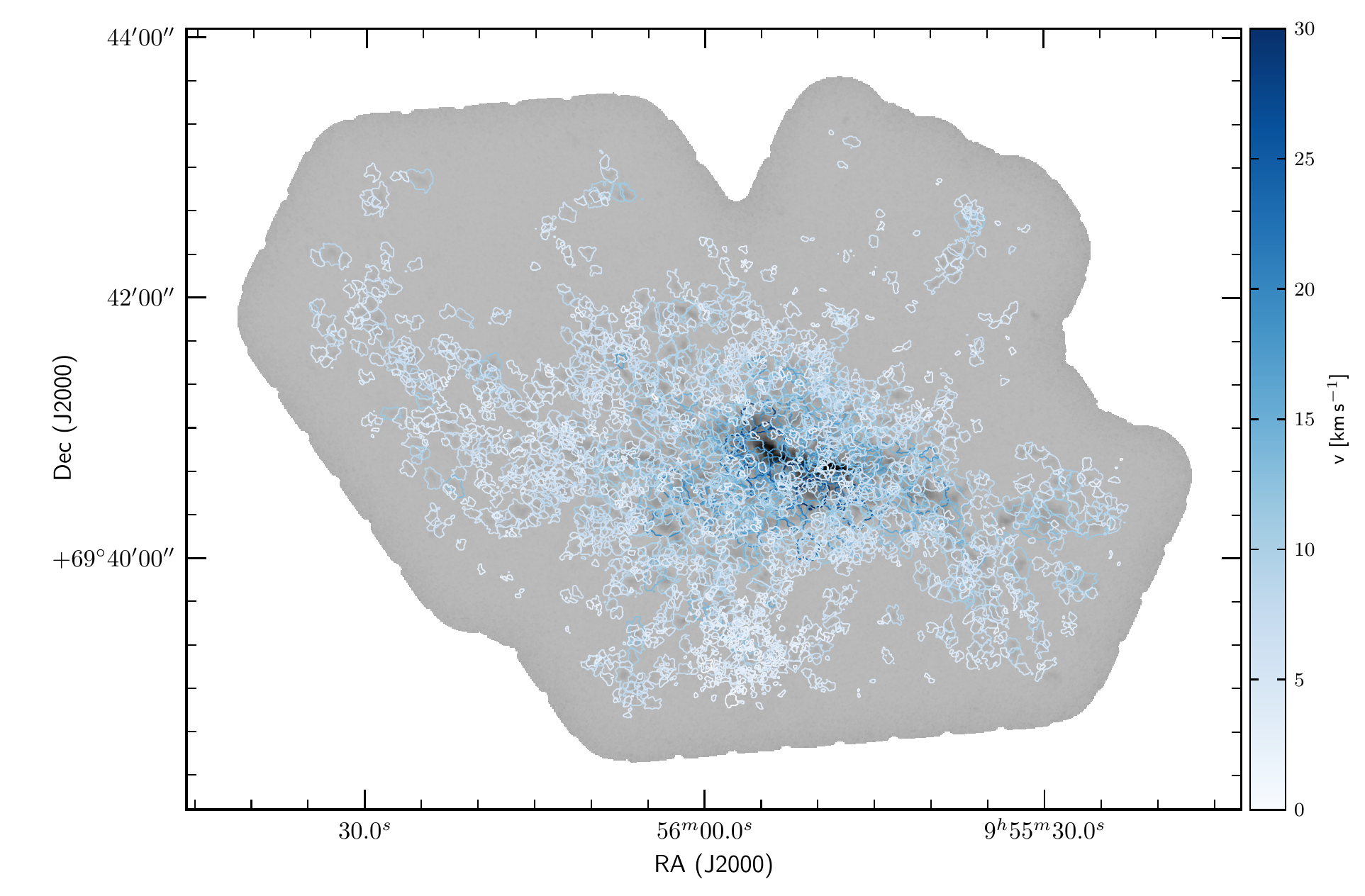}
    \caption{Contours for all 1891 clouds identified in Sec.~\ref{section: Fellwalker}, color-coded by systemic velocity (\emph{left}) and line width (\emph{right}), overplotted on the \co10 peak intensity map.
    \label{figure: cloud contours}}
\end{figure*}

The output of Fellwalker is a cloud assignment array that labels each pixel with the ID of the cloud that it belongs to. In the left panel of Fig.~\ref{figure: cloud contours} we show the 1891 clouds obtained by Fellwalker and project them onto the \co10 peak intensity map, color--coded by their respective systemic velocity. The projection results in many clouds that are situated on top of each other. The superposition of these overlapping clouds then leads to the high dispersion measurements discussed in Sec.~\ref{section: pixel statistics}, but the individual clouds often have narrower line widths. We indicate the velocity dispersions (Sec.~\ref{section: cloud dispersion} of individual clouds as colors in the right panel of Fig.~\ref{figure: cloud contours}, again overplotted on top of the \co10 peak intensity map.

The cloud catalog is available in machine-readable format (Appendix~\ref{section: cloud catalog}, Tab.~\ref{table: cloud catalog}).

\begin{figure}
    \centering
    \includegraphics[width=\linewidth]{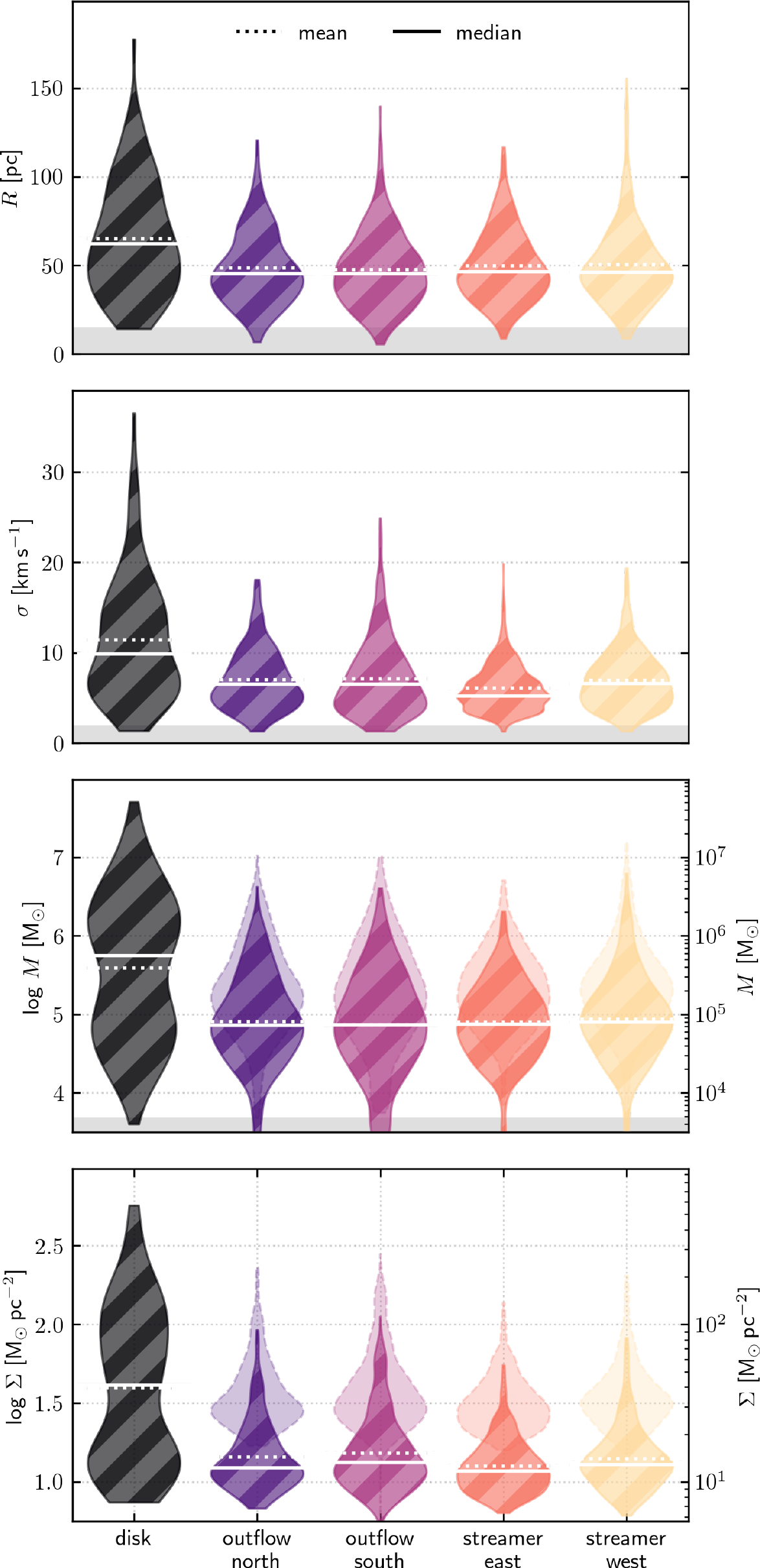}
    \caption{Distributions of radius $R$, velocity dispersion $\sigma$, mass $M$ and surface density $\Sigma$ compared across the five regions (Sec.~\ref{section: region definition}). Each violin is a histogram along the vertical axis. Grey bands for radius, velocity dispersion and mass show the resolution limits. The lightly colored violins with dashed contours for mass and surface density assume a higher conversion factor ($\alpha_\mathrm{CO} = 2.5$\,\alphaCOunit) applicable to faint clouds instead of a starburst disk conversion factor ($\alpha_\mathrm{CO} = 1.0$\,\alphaCOunit, hatched violins) as suggested by \citet{2015ApJ...814...83L}.
    \label{figure: violins}}
\end{figure}

\subsection{Cloud Property Measurements}\label{section: cloud measurements}

\subsubsection{Cloud Radii}\label{section: cloud radius}

We start by using the Fellwalker output to calculate the radius for each cloud. We measure the circularize radius, $R$ from $R = \sqrt{A}/\pi$ where $A$ is the projected two-dimensional area of the pixels in the Fellwalker-generated cloud assignment.
As we discussed in detail in Appendix~A of \citet{2020ApJ...899..158K}, this definition of radius differs by a constant factor from alternative definitions when considering cloud ensembles. Note that for individual clouds the picture is  more complex and needs to take the internal cloud structure into account. These effects, however, are below the resolution limit of this study. Therefore, the choice of radius definition does not affect the comparative study of statistical properties presented here. 
To account for our instrumental resolution, we deconvolve the radius by subtracting the half width--half maximum beam size in quadrature. Note that this represents an approximation, because the area of a cloud in the data cube is a function of both the cloud size and signal-to-noise; e.g., a cloud with peak signal-to-noise of $50$ has a larger footprint than an otherwise identical cloud with peak signal-to-noise of $3$. The FWHM beam size used in convolution would be appropriate for peak $\mathrm{SNR} = 2$ and so should be approximately correct for the faint clouds in the extended streamers and outflows.

We show the resulting distributions of radius in the top left panel in Fig.~\ref{figure: violins}. Most of our clouds have radii in the range of $40-60$\,pc, well above the observational limit. Larger clouds are found in the disk (median $\sim 60$\,pc, with values up to $\sim 150$\,pc) compared to the outflows and streamers (medians $40-45$\,pc). There appears to be no statistical difference in the radii of an average cloud found in the outflow compared to those in the streamers. 
We note that by-chance superpositions of clouds are more likely in edge-on disks and may partially contribute to the difference between disk and outflows or streamers.

\subsubsection{Cloud Dispersions}\label{section: cloud dispersion}

We also measure the line width, or rms velocity dispersion, of each cloud. To do this, we calculate the intensity-weighted second moment for each pixel in the cloud. Then we take the characteristic rms velocity to be the median of this pixel-by-pixel intensity--weighted second moment map over all pixels in the cloud.
Similar to the radius, other definitions of line width differ for ensembles of clouds, from the one used here only by constant factors (Appendix~A in \citealt{2020ApJ...899..158K}).
We correct the line widths for the effect of instrumental resolution by deconvolving the measured width with the Gaussian equivalent of the instrumental top hat profile. Formally, the rms velocity dispersion of a top hat channel profile is $\delta v/\sqrt{2 \pi}$, so our correction is slightly conservative.

In Fig.~\ref{figure: violins} (second from top), we show the distribution of the rms velocity dispersion $\sigma$ of individual clouds. The median dispersion differs between the disk ($\sim 9.8$\,\kms) and the other regions ($5.2-6.5$\,\kms) but not so much between the outflows (north: median 6.0\,\kms, south: median 5.2\,\kms) and the streamers (east: median 5.2\,\kms, west: median 6.5\,\kms). In all regions, a value of 5-8\,\kms is the most common velocity dispersion, however the disk has a wider distribution with prevalent values up to $\sim 20$\,\kms. 
As for radii, also the cloud velocity dispersions in the disk are more likely to be enhanced by chance superpositions than in the outflows or streamers.

Compared to the numbers derived from the second moment map (Sec.~\ref{section: pixel statistics}), the dispersion values calculated from the individual clouds identified by Fellwalker are significantly smaller. This reflects the fact that calculating second moment measurements based on the full data cube, overlapping clouds will lead to higher line-of-sight velocity dispersions.

We note that our choices for the derivations of cloud radius and velocity dispersion are conceptionally different (based on total area vs. based on intensity weighted dispersion). This is due to the need for robust methods that work across a wide range of sizes, SNR and environment. However, a statistical comparison of cloud properties is largely insensitive to the details how cloud properties are defined.

\subsubsection{Cloud Luminosities and Masses}\label{section: cloud masses}

We calculate the integrated flux of each cloud by summing the CO flux inside the cloud assignment created by Fellwalker. Combining this flux with the adopted distance to M82, we calculate a \co10 luminosity for each cloud.

We also make a rough estimate of the mass of each cloud by scaling its CO luminosity. This requires assuming a CO-to-H$_2$ conversion factor. This factor is uncertain, and so we adopt two approaches that likely span the range of possible values. First, we use a common conversion factor of $\alphaCO =$ 1.0\,\alphaCOunit for all regions. This is the value for the starburst disk and the bright regions of the outflow derived by \citet{2015ApJ...814...83L} comparing dust and CO emission. This estimate is consistent with the typical starburst value ($\alphaCO = 0.8$\,\alphaCOunit, \citealt{2013ARA&A..51..207B}). The masses derived applying this starburst conversion factor to the whole map appear as hatched, dark violins in Figure \ref{figure: violins}. 
We might expect the CO-to-H$_2$ conversion factor to vary across the galaxy. As the conversion factor in the streamers and the outflow is unknown, we also calculate a version of the masses using a higher conversion factor of $\alphaCO = 2.5$\,\alphaCOunit as proposed by \citet{2015ApJ...814...83L} for these  regions. The resulting numbers are shown as dashed, lightly colored violins in Fig.~\ref{figure: violins}.

The distribution of luminosity-based cloud masses for each region appears in Fig.~\ref{figure: violins}. In the disk, we find median values of $5.8\times10^5$\,\Msun. In the outflows and streamers we find cloud masses that are significantly lower (median $5.5-7.6 \times 10^4$\,\Msun) than in the disk even when considering the case where these regions have a higher conversion factor (dashed violins, medians $1.4-1.9 \times 10^5$\,\Msun). As was the case with the radius and the velocity dispersion, there do not appear to be any significant statistical differences between the mass distribution in the outflows and that in the streamers.

The total gas mass of all clouds identified by Fellwalker is $6.19\times10^8$\,\Msun, assuming a starburst conversion factor ($\alphaCO = 1.0$\,\alphaCOunit) throughout, and $1.09\times 10^9$\,\Msun assuming two different conversion factors, as discussed above. 
This is $\sim 75$\% of the total mass obtained from the moment 0 map and slightly lower but consistent with previous interferometric imaging \citep[\co10,][]{2002ApJ...580L..21W} and agrees well with single dish observations (e.g. \citealt{2013PASJ...65...66S} for \co10 or \citealt{2015ApJ...814...83L} for \co21). In those studies, the masses were based on the integrated maps and not individual clouds.
The emission that is not recovered in clouds is located in the outflow and streamers and is coincident with structures that are below the cloud limits (see Sec.~\ref{section: Fellwalker}). The bright disk is recovered completely by the cloud decomposition.
The recovery fraction in the low SNR regime could only be increased by lowering the detection thresholds. However, this would in turn lead to an increasing number of noise peaks being detected as clouds which would affect the cloud statistics, which we aim to avoid.

We calculate the cloud surface density $\Sigma = M/A$, which is the surface brightness scaled by a conversion factor, and show the resulting distributions in the bottom panel of Fig.~\ref{figure: violins}.
Following from cloud mass, surface density is also derived from luminosity and depends on the adopted CO-to-H$_2$ conversion factor in the same way as is discussed above.
We further note that surface density is less dependent on observational limits than radius or mass because they partially divide out in the calculation.

We find typical surface densities of $12-140$\,\Msunpc2 in the disk (median 42\,\Msunpc2). In the outflows and streamers, the inferred surface densities fall in the range $8-25$\,\Msunpc2 (medians $11-13$\,\Msunpc2). As for the previously discussed parameters, we again find no differences between the outflows and the streamers.

\subsection{Galactocentric Distance Dependence of Cloud Properties}\label{section: cloud radial dependence}

\begin{figure}
    \centering
    \includegraphics[width=\linewidth]{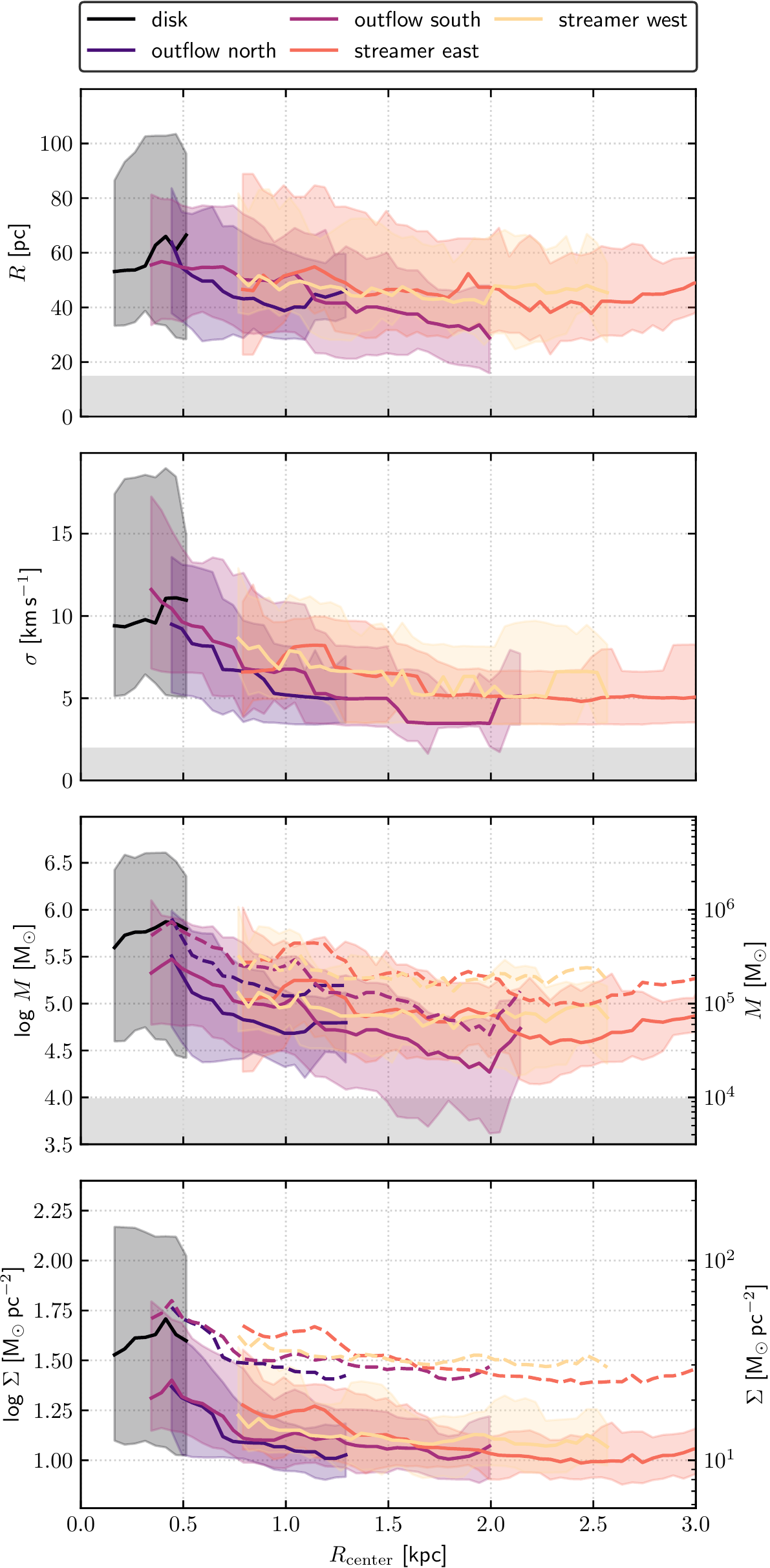}
    \caption{Dependence of radius, velocity dispersion, mass and surface density on the distance from \m82's center. Shown is the sliding median over bins of 0.1\,kpc width together with the 16$^\mathrm{th}$ to 84$^\mathrm{th}$ percentile of the distribution. Horizontal grey bands at the bottom fo the radius, velocity dispersion and mass panels show the resolution limits. For the quantities that depend on a conversion factor, two choices for the starburst disk and bright clouds ($\alphaCO = 1.0$, continuous) and faint clouds ($\alphaCO = 2.5$, dashed) are shown.
    \label{figure: radial dependence}}
\end{figure}

We now explore the dependence of the cloud properties on projected distance from the galaxy center. To do so, we define the projected galactocentric radius as its distance to \m82's center at ($\alpha, \delta$) = ($09^h 55^m 52.72^s$, $69^\circ 40\arcmin 45.7\arcsec$). Note that we make no correction for inclination or orientation, our ``radius'' is physical distance along the plane of the sky.

Fig.~\ref{figure: radial dependence} shows the radial trends of each quantity discussed above using sliding medians over bins with a width of 0.1\,kpc.
As the disk is viewed in edge-on projection and is highly asymmetric, the change of disk cloud properties as a function of distance is not meaningful using this radius definition, and is not discussed further. We suggest to view the disk properties primarily as interesting in their contrast with the properties in the other regions.
However, it is interesting to look at the trends both for the outflows and the streamers. 

\emph{Radii:} We find that the radii for the clouds in the streamers stay approximately constant as a  function of radii. The radii for the clouds in the outflow, in particular towards the south, however appear to decrease with increasing distance.

\emph{Dispersions:} The radial trends in $\sigma$ show that, on average, the velocity dispersion decreases with increasing distance. The effect is strongest in the outflow over the range $0.5<R_\mathrm{center}<1.5$\,kpc, where the line width appears to drop by almost a factor of $2$ from $\sim 10$\,\kms to 5\,\kms.  At larger distance, in the streamers, the gradient becomes much shallower or $\sigma$ stays approximately constant.

\emph{Masses:} In the case of the streamers, the individual cloud masses stay approximately constant, or show at most a mild decline, as a  function of distance. The masses of the clouds in both the southern and northern outflows show a clear, rapid decrease with radius.
It is unknown how the conversion factor \alphaCO, as the primary source of uncertainty, varies with galactocentric distance. It could be expected that \alphaCO increases with distance from the disk value to the outflow/streamer value proposed by \citet{2015ApJ...814...83L}. In this case, the masses in the streamers would, to first order, be constant with distance. The cloud masses in the outflows would still decline with distance, but with a more shallow slope.

\emph{Surface density:} The surface density in the outflows drops steeply by a factor of $\sim 2$ within only $\Delta R_\mathrm{center} \sim 0.5$\,kpc and stays approximately constant beyond $\sim 1$\,kpc. In the streamers, the surface density is only slowly decreasing with distance.
Taking the uncertainty in the conversion factor into account would allow for a constant or mildly increasing surface density as a function of distance in both the outflows and the streamers. If the conversion factor were to vary in a clumpy manner instead of a smooth gradient, more complex radial profiles of surface density may be possible.

\subsection{Physical State of the Clouds}\label{section: cloud virialization}

We now compare the physical state of the individual molecular clouds in the different environments probed by the observations. We do so by first looking at the size--line width relation in Sec.~\ref{section: cloud radius linewidth}, followed by assessing the effects of external pressure between the different environments in Sec.~\ref{section: pressure}.

\subsubsection{Size -- Line Width Relation}\label{section: cloud radius linewidth}

\begin{figure}
    \centering
    \includegraphics[width=\linewidth]{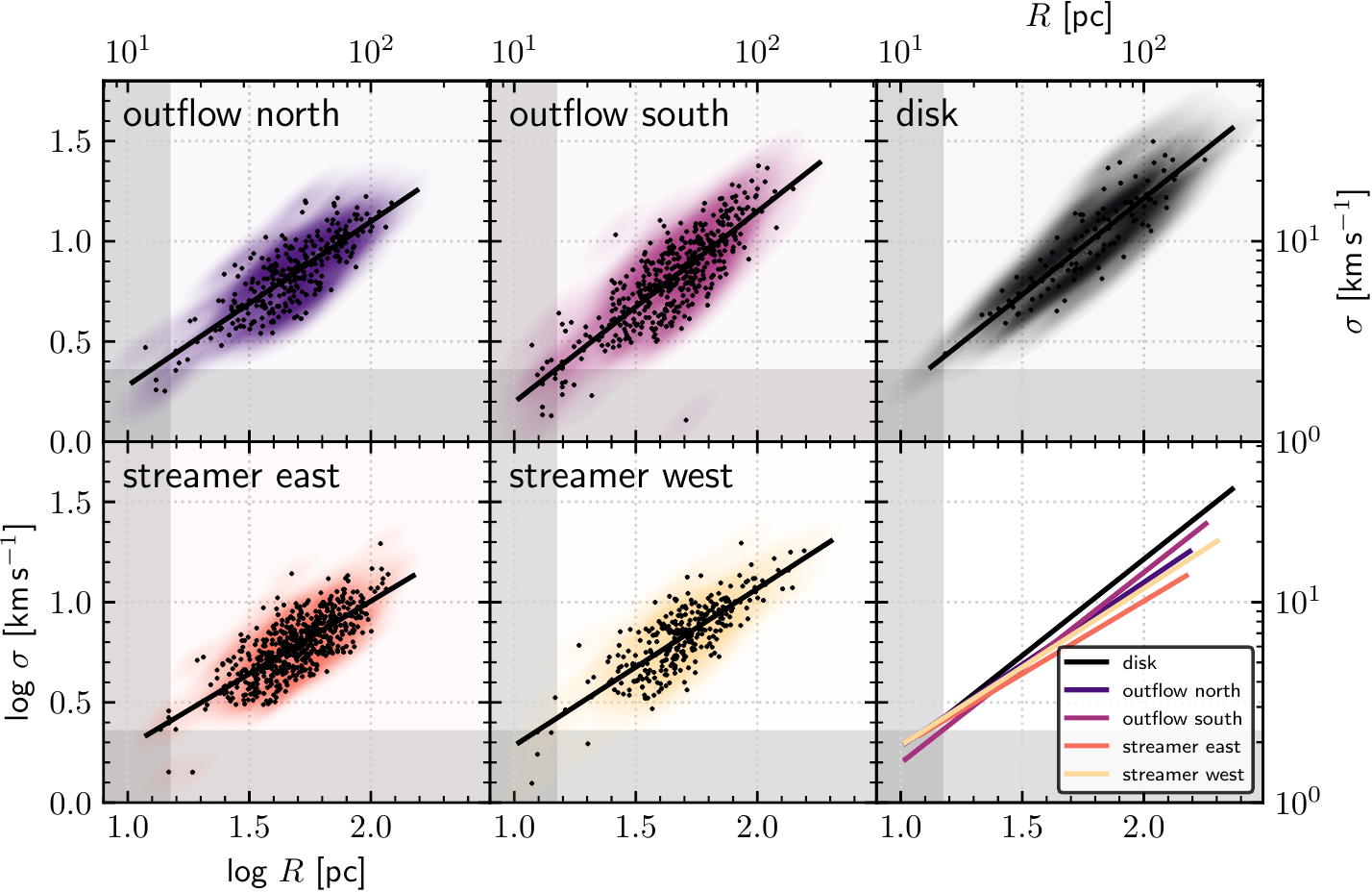}
    \caption{Size -- line width relation of the clouds, comparing between disk, outflow and tidal streamers. Each panel shows all clouds in that region (black dots), their density (colored) and a power law fit (black line) as a representative line that follows the distribution. Grey bands show the resolution limits. The bottom right panel compares the lines and allows to relate the panels relative to each other.
    \label{figure: size line width}}
\end{figure}

\begin{deluxetable}{lCCC}
	\tablewidth{\linewidth}
	\tablecaption{Fits to the size -- line width relations $\sigma = aR^b = 10^{R_{10}}R^b$.
	\label{table: size line width fits}}
	\tablehead{\colhead{region} & \colhead{a} & \colhead{$R_{10}$} & \colhead{b}}
	\startdata
disk          & 0.20 & 1.57 & 0.96\\
outflow north & 0.29 & 1.95 & 0.82\\
outflow south & 0.18 & 1.50 & 0.95\\
streamer east & 0.36 & 2.28 & 0.73\\
streamer west & 0.32 & 2.07 & 0.78\\
	\enddata
    \tablecomments{$R_{10}$ is the line width at a representative size scale of 10\,pc. The statistical fitting errors are negligible and dominated by the systematic errors.
    We note that these values should only be used for a relative comparison of the five regions presented in this study.}
\end{deluxetable}

In a turbulent medium, the measured line width will be larger when considering a larger size scale \citep[e.g.,][]{Larson:1981jma}. To compare the different regions independent of this overall scaling, Fig.~\ref{figure: size line width} shows the size (radius) -- line width (rms velocity dispersion) relation for each of our five regions.
In all regions, the line widths of the clouds scale with radius, suggestive of a turbulent medium, and follow a power law relation (Tab.~\ref{table: size line width fits}).
As shown in the last panel and Tab.~\ref{table: size line width fits}, the relations between the outflow and the streamer regions appear indistinguishable within the uncertainties. Moreover, despite finding larger variation of $R$ and $\sigma$ between the disk and other regions, the size -- line width relation in the disk is also very similar to that of the outflow and streamers. 

In other words, the bottom right panel in Fig.~\ref{figure: size line width} shows that the clouds we identify in \m82 appear to populate a common size--line width relation regardless of region. The primary difference between the regions appears to be what part of the relation is populated. The disk clouds show high $R$ and high $\sigma$, whereas the clouds in the tidal streamers and the outflows populate the comparatively low $R$ and low $\sigma$.

\subsubsection{Size, Line Width, and Surface Brightness or Surface Density}
\label{section: pressure}

\begin{figure}
    \centering
    \includegraphics[width=\linewidth]{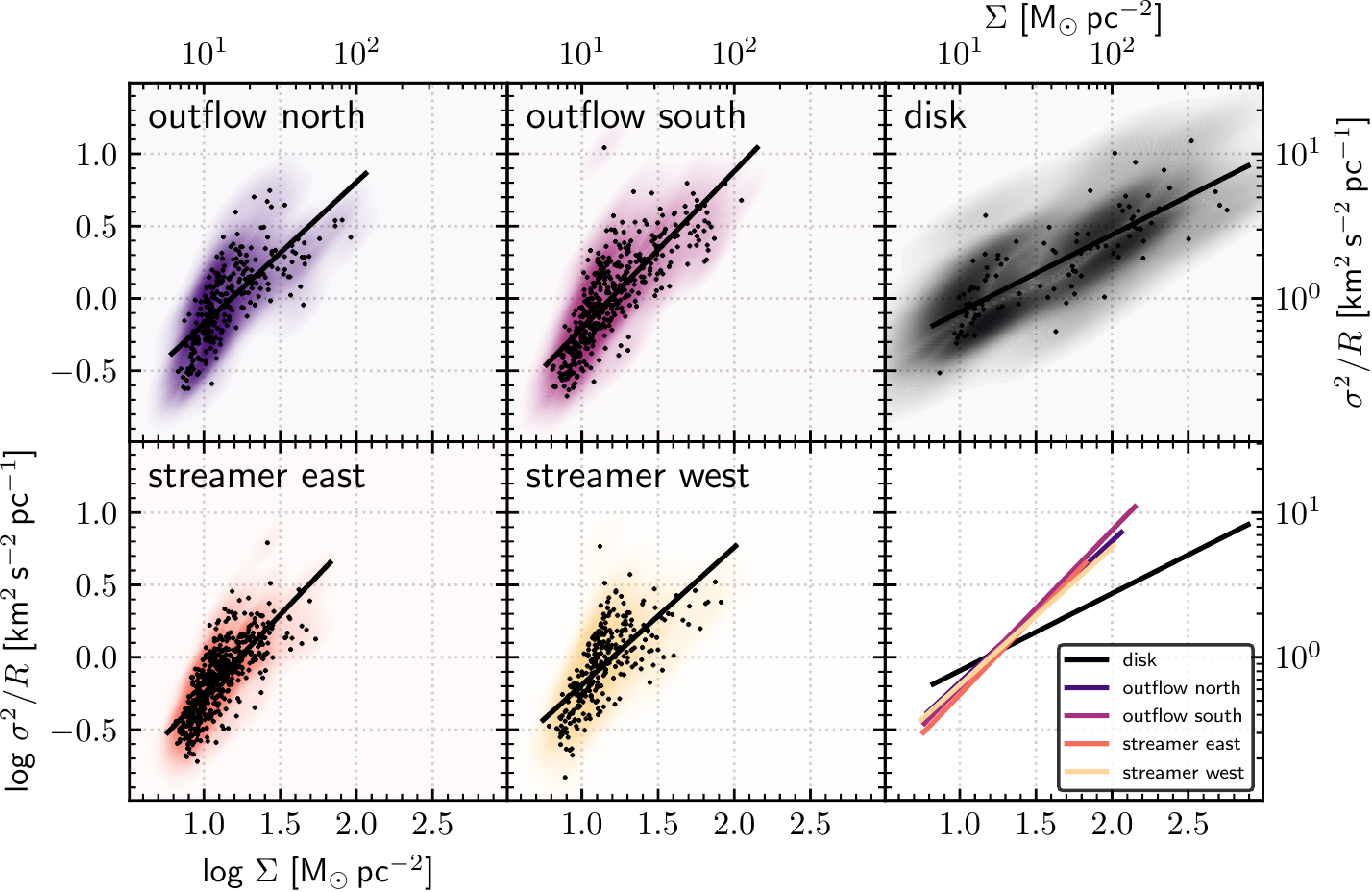}
    \caption{Relationship between cloud surface density $\Sigma$ and size -- line width coefficient $\sigma^2/R$ in different regions of \m82. Each panel shows all clouds in this region (black dots), their density (colored) and a power law fit (black line) as a representative line that follows the distribution. The bottom right panel compares the lines and allows to relate the panels relative to each other.
    \label{figure: density coeff}}
\end{figure}

For a self-gravitating cloud, meaning a marginally bound cloud, a cloud in virial equilibrium, or even a free-falling cloud, we expect a relationship between size, line width, and mass surface density such that $\sigma^2/R \propto \Sigma$ \citep[e.g.][]{1986ApJ...304..466K,2011MNRAS.416..710F}. Deviations from this relationship can give insight into the dynamical state of clouds, or they can highlight uncertainties with the physical parameter estimation. Although the mass, and so surface density, estimates for our clouds are uncertain, we can still gain insight into the dynamic state of the clouds and the origins of the observed line widths. We plot $\sigma^2/R$ vs. $\Sigma$ in Fig.~\ref{figure: density coeff}.

The figure shows significant scatter but a similar scaling between $\Sigma$ and $\sigma^2/R$ for each of the outflow and streamer regions. The outflow regions tend to show more high surface brightness and high line width points than the streamers, reflecting mostly bright emission near the disk. In all four regions the slope is slightly steeper than $\sim 1$ but the data appear consistent with a line of slope $\sim 1$. We do not trust our adopted conversion factor enough to know if the clouds are bound, virialized, or in another state. But their combined line width-radius-surface density scaling is consistent with self-gravity playing an important role.

The situation appears different in the central disk, where we observe a shallower slope and a very wide range of surface brightness or surface density. This could reflect that the bright, dense regions are self-gravitating, while the lower surface density regions form part of a more extended, diffuse molecular medium. In this case, the lower surface brightness clouds might be more affected by ``external pressure'' or simply more dominated by turbulent motions, representing temporary fluctuations in a turbulent medium. Alternatively, it might indicate significant changes in the conversion factor across the disk region. In any case, the differences may be influenced by chance superpositions of clouds that are geometrically more likely in the (close to) edge-on disk.

\section{Summary and Conclusion}\label{section: summary}

We present a 154 pointing NOEMA mosaic of the \co10 line emission in and around the nearby starburst galaxy \m82. The observations reach a spatial resolution of $\sim$30\,pc, sufficient to spatially resolve the molecular gas in the central starburst disk, the molecular outflow, as well the tidal streamers. We obtained a striking \co10 data cube in which we have identified 1891 individual molecular clouds. These data  enable us to study the properties of the molecular gas in greatly distinct environments, in particular:

\emph{Tidal features:} also known as `streamers' that are thought to be due to a recent interaction with \m81. The molecular features coincide with the tidal features seen in \ion{H}{1} emission \citep[e.g.][]{1993ApJ...411L..17Y,2018ApJ...865...26D}, and are not thought to harbor any significant star formation activity. The expectation is that these regions are `cold', i.e. not highly excited by an external radiation field.

\emph{Outflows:} molecular gas is clearly associated with \m82's prominent outflow. This outflow is known to harbor hot X--ray gas, with temperatures exceeding 10$^6$\,K \citep[e.g.][]{1999ApJ...523..575L,2008MNRAS.386.1464R,2020ApJ...904..152L}. Even though it is currently unknown how the molecular gas is associated within the hot outflow (mixed gas vs.\ entrained emission), the environment of the gas in the outflow is quite different from that of the cold tidal features.

\emph{Disk:} the molecular gas in \m82's disk has been the study of many investigations, and is not the main target of this study. This environment is known to be very extreme, yielding the high star formation rate density in the central disk that gives rise to the prominent outflow. 

These observations thus allow us to compare molecular cloud properties in distinctly different environments. In particular, the molecular outflow and the tidal streamers have never been observed at such high spatial and spectral resolution. This allows us to separate clouds that are close in (projected) position, but have significant offsets in velocities. In the tidal arms this indicates overlapping velocity components due to previous galaxy interactions. In the outflows, these offsets are likely due to projection effects of the front and backside of the outflow cone.

There are tendencies for the clouds in the outflow to have smaller sizes ($\sim 0.2$\,dex\,kpc$^{-1}$), lower velocity dispersions ($\sim 0.3-0.4$\,dex\,kpc$^{-1}$), lower masses (computed from their luminosity, $\sim 0.7$\,dex\,kpc$^{-1}$) and lower surface densities (surface brightness, $\sim 0.3-0.4$\,dex\,kpc$^{-1}$) at larger distances from the galaxy (Figure \ref{figure: radial dependence}). The decrease in size, dispersion and mass (luminosity) is particularly clear towards the more prominent southern outflow. The reduction in the molecular gas surface density (surface brightness) is consistent with that observed in previous work at significantly lower spatial resolution \citep{2015ApJ...814...83L}.
Together, these trends could be indicative of evaporation of the clouds embedded in the hot outflow.
To further study the fate of the molecular clouds, multi--line high resolution and sensitivity measurements will be needed alongside simulations that can be directly compared to the observations.

Unlike the clouds in the outflow, we find that the clouds in the streamers stay approximately constant in size and mass, irrespective of distance to \m82. They slowly decrease in line width ($\sim 0.1$\,dex\,kpc$^{-1}$) and surface density ($\sim 0.1$\,dex\,kpc$^{-1}$). Also, the  size--linewidth relations for the clouds show indistinguishable behavior between the extra--galaxy regions and the galaxy disk itself (Fig.~\ref{figure: size line width}). The distribution of individual clouds in the $\sigma^2/R$ vs.\ $\Sigma$ space does not show obvious effects due to changing external pressure in the outflow or streamer clouds, although the  disk clouds show a different behavior (Fig.~\ref{figure: density coeff}.

This is a first analysis of this extremely rich and complex dataset of \m82. In the future, we expect to include $^{13}$CO information to have an independent estimator of surface density that would enable us to more thoroughly assess the dynamical state of the clouds in the outflow and streamers, as well as multi--transition CO spectroscopy. These data will enable comparisons with increasingly realistic simulations of starburst--driven, multi--phase outflows in galaxies \citep[e.g.][]{2020MNRAS.499.4261S,2020ApJ...903L..34K}.


\acknowledgements

We thank the referee for a very constructive report that helped to improve the analysis presented in this paper. This work is based on observations carried out under project number w18by and 107-19 with the IRAM NOEMA Interferometer and the IRAM 30\,m telescope, respectively. IRAM is supported by INSU/CNRS (France), MPG (Germany) and IGN (Spain).

\facilities{IRAM NOEMA, IRAM 30\,m}
\software{Gildas \footnote{\url{https://www.iram.fr/IRAMFR/GILDAS}}, CASA \citep{McMullin:2007tj}, Astropy \citep{Collaboration:2013cd,Collaboration:2018ji}, NumPy \citep{Harris:2020fx}, SciPy \citep{2020NatMe..17..261V}, pyCupid\footnote{\url{https://pycupid.readthedocs.io/en/latest/index.html}}, spectral-cube\footnote{\url{https://spectral-cube.readthedocs.io}}}

\clearpage
\appendix

\section{Cloud catalog}\label{section: cloud catalog}

The cloud catalog derived in Sec.~\ref{section: discussion} is available in the machine-readable format. It includes position, region attribution, distance from \m82's center, size, line width, as well as mass and surface density derived from surface brightness for the 1891 clouds. Tab.~\ref{table: cloud catalog} shows an abbreviated overview.

Please note that comparisons to other datasets using different methodologies or different targets need to be considered carefully. The definitions and parameters (e.g. size, line width or conversion factor) used here may not match other works, which can lead to differences in normalization or scaling of cloud properties.

\begin{deluxetable*}{CCCCcCCCCCC}
\tablecaption{Cloud catalog
    \label{table: cloud catalog}
}
\tablehead{\colhead{index} & \colhead{RA} & \colhead{DEC} & \colhead{V} & \colhead{region} & \colhead{$R_\mathrm{center}$} & \colhead{$R$} & \colhead{$\sigma$} & \colhead{$M$} & \colhead{$\Sigma$} & \colhead{$\mu$}\\
\colhead{ } & \colhead{$\mathrm{{}^{\circ}}$} & \colhead{$\mathrm{{}^{\circ}}$} & \colhead{$\mathrm{km\,s^{-1}}$} & \colhead{ } & \colhead{$\mathrm{kpc}$} & \colhead{$\mathrm{pc}$} & \colhead{$\mathrm{km\,s^{-1}}$} & \colhead{$\mathrm{M_{\odot}}$} & \colhead{$\mathrm{M_{\odot}\,pc^{-2}}$} & \colhead{$\mathrm{K\,km\,s^{-1}\,pc^{-2}}$}\\
 & \colhead{(a)} & \colhead{(b)} & \colhead{(c)} & \colhead{(d)} & \colhead{(e)} & \colhead{(f)} & \colhead{(g)} & \colhead{(h)} & \colhead{(i)} & \colhead{(j)}
}
\startdata
1 & 148.955 & 69.677 & 110.0 & disk & 0.32 & 134.7 & 23.5 & 3.23 \times 10^7 & 568.57 & 11.49 \\
2 & 148.971 & 69.680 & 250.2 & disk & 0.04 & 122.6 & 17.8 & 1.50 \times 10^7 & 317.91 & 6.40 \\
3 & 148.975 & 69.681 & 315.3 & disk & 0.16 & 177.5 & 28.0 & 5.03 \times 10^7 & 508.83 & 10.35 \\
... & ... & ... & ... & ... & ... & ... & ... & ... & ... & ...\\
1889 & 148.948 & 69.688 & 160.1 & outflow north & 0.73 & 16.2 & 3.3 & 9.22 \times 10^3 & 11.21 & 0.114 \\
1890 & 148.980 & 69.660 & 85.0 & outflow south & 1.24 & 19.6 & 3.5 & 1.40 \times 10^4 & 11.55 & 0.140 \\
1891 & 148.934 & 69.669 & 130.0 & streamer west & 1.02 & 49.2 & 12.6 & 1.07 \times 10^5 & 14.13 & 0.261\\
\enddata
\tablecomments{The cloud catalog is published in its entirety in the machine-readable format. A portion is shown here for guidance regarding its form and content.\\
(a-c) Location of the peak within the cloud. (d) Region according to the definition in Sec.~\ref{section: region definition}. (e) Distance to \m82's center at ($\alpha, \delta$) = ($09^h 55^m 52.72^s$, $69^\circ 40\arcmin 45.7\arcsec$). (f) Cloud radius according to Sec.~\ref{section: cloud radius}. (g) Cloud velocity dispersion according to Sec.~\ref{section: cloud dispersion}. (h) Cloud mass assuming a starburst conversion factor of $\alphaCO =$ 1.0\,\alphaCOunit. (i) Cloud surface density assuming a starburst conversion factor of $\alphaCO =$ 1.0\,\alphaCOunit. (j) Cloud surface brightness.
}
\end{deluxetable*}

\clearpage
\bibliography{bibliography.bib}{}

\begin{thebibliography}{}
\expandafter\ifx\csname natexlab\endcsname\relax\def\natexlab#1{#1}\fi
\providecommand{\url}[1]{\href{#1}{#1}}
\providecommand{\dodoi}[1]{doi:~\href{http://doi.org/#1}{\nolinkurl{#1}}}
\providecommand{\doeprint}[1]{\href{http://ascl.net/#1}{\nolinkurl{http://ascl.net/#1}}}
\providecommand{\doarXiv}[1]{\href{https://arxiv.org/abs/#1}{\nolinkurl{https://arxiv.org/abs/#1}}}

\bibitem[{Abruzzo {et~al.}(2021)Abruzzo, Bryan, \& Fielding}]{Abruzzo:2021uv}
Abruzzo, M.~W., Bryan, G.~L., \& Fielding, D.~B. 2021, arXiv.org

\bibitem[{Beir{\~a}o {et~al.}(2015)Beir{\~a}o, Armus, Lehnert, Guillard,
  Heckman, Draine, Hollenbach, Walter, Sheth, Smith, Shopbell, Boulanger,
  Surace, Hoopes, \& Engelbracht}]{2015MNRAS.451.2640B}
Beir{\~a}o, P., Armus, L., Lehnert, M.~D., {et~al.} 2015, Monthly Notices of
  the Royal Astronomical Society, 451, 2640

\bibitem[{Berry(2014)}]{Berry:2014ha}
Berry, D. 2014, arXiv.org, 22

\bibitem[{Bland \& Tully(1988)}]{1988Natur.334...43B}
Bland, J., \& Tully, B. 1988, Nature (ISSN 0028-0836), 334, 43

\bibitem[{Bolatto {et~al.}(2008)Bolatto, Leroy, Rosolowsky, Walter, \&
  Blitz}]{Bolatto:2008iv}
Bolatto, A.~D., Leroy, A.~K., Rosolowsky, E., Walter, F., \& Blitz, L. 2008,
  The Astrophysical Journal, 686, 948

\bibitem[{Bolatto {et~al.}(2013{\natexlab{a}})Bolatto, Wolfire, \&
  Leroy}]{2013ARA&A..51..207B}
Bolatto, A.~D., Wolfire, M., \& Leroy, A.~K. 2013{\natexlab{a}}, Annual Review
  of Astronomy and Astrophysics, 51, 207

\bibitem[{Bolatto {et~al.}(2013{\natexlab{b}})Bolatto, Warren, Leroy, Walter,
  Veilleux, Ostriker, Ott, Zwaan, Fisher, Wei{\ss}, Rosolowsky, \&
  Hodge}]{2013Natur.499..450B}
Bolatto, A.~D., Warren, S.~R., Leroy, A.~K., {et~al.} 2013{\natexlab{b}},
  Nature, 499, 450

\bibitem[{Chisholm \& Matsushita(2016)}]{2016ApJ...830...72C}
Chisholm, J., \& Matsushita, S. 2016, The Astrophysical Journal, 830, 72

\bibitem[{Collaboration {et~al.}(2013)Collaboration, Robitaille, Tollerud,
  Greenfield, Droettboom, Bray, Aldcroft, Davis, Ginsburg, Price-Whelan,
  Kerzendorf, Conley, Crighton, Barbary, Muna, Ferguson, Grollier, Parikh,
  Nair, Unther, Deil, Woillez, Conseil, Kramer, Turner, Singer, Fox, Weaver,
  Zabalza, Edwards, Azalee~Bostroem, Burke, Casey, Crawford, Dencheva, Ely,
  Jenness, Labrie, Lim, Pierfederici, Pontzen, Ptak, Refsdal, Servillat, \&
  Streicher}]{Collaboration:2013cd}
Collaboration, A., Robitaille, T.~P., Tollerud, E.~J., {et~al.} 2013, Astronomy
  {\&} Astrophysics, 558, A33

\bibitem[{Collaboration {et~al.}(2018)Collaboration, Price-Whelan, Sip{\H o}cz,
  G{\"u}nther, Lim, Crawford, Conseil, Shupe, Craig, Dencheva, Ginsburg,
  VanderPlas, Bradley, P{\'e}rez-Su{\'a}rez, de~Val-Borro, Aldcroft, Cruz,
  Robitaille, Tollerud, Ardelean, Babej, Bach, Bachetti, Bakanov, Bamford,
  Barentsen, Barmby, Baumbach, Berry, Biscani, Boquien, Bostroem, Bouma,
  Brammer, Bray, Breytenbach, Buddelmeijer, Burke, Calderone,
  Cano~Rodr{\'\i}guez, Cara, Cardoso, Cheedella, Copin, Corrales, Crichton,
  D{\'A}vella, Deil, Depagne, Dietrich, Donath, Droettboom, Earl, Erben,
  Fabbro, Ferreira, Finethy, Fox, Garrison, Gibbons, Goldstein, Gommers, Greco,
  Greenfield, Groener, Grollier, Hagen, Hirst, Homeier, Horton, Hosseinzadeh,
  Hu, Hunkeler, Ivezi{\'c}, Jain, Jenness, Kanarek, Kendrew, Kern, Kerzendorf,
  Khvalko, King, Kirkby, Kulkarni, Kumar, Lee, Lenz, Littlefair, Ma, Macleod,
  Mastropietro, McCully, Montagnac, Morris, Mueller, Mumford, Muna, Murphy,
  Nelson, Nguyen, Ninan, N{\"o}the, Ogaz, Oh, Parejko, Parley, Pascual, Patil,
  Patil, Plunkett, Prochaska, Rastogi, Reddy~Janga, Sabater, Sakurikar,
  Seifert, Sherbert, Sherwood-Taylor, Shih, Sick, Silbiger, Singanamalla,
  {Singer, L. P.}, Sladen, Sooley, Sornarajah, Streicher, Teuben, Thomas,
  Tremblay, Turner, Terr{\'o}n, van Kerkwijk, de~la Vega, Watkins, Weaver,
  Whitmore, Woillez, Zabalza, \& Contributors}]{Collaboration:2018ji}
Collaboration, A., Price-Whelan, A.~M., Sip{\H o}cz, B.~M., {et~al.} 2018, The
  Astronomical Journal, 156, 123

\bibitem[{de~Blok {et~al.}(2018)de~Blok, Walter, Ferguson, Bernard, van~der
  Hulst, Neeleman, Leroy, Ott, Zschaechner, Zwaan, Yun, Langston, \&
  Keating}]{2018ApJ...865...26D}
de~Blok, W. J.~G., Walter, F., Ferguson, A. M.~N., {et~al.} 2018, The
  Astrophysical Journal, 865, 26

\bibitem[{Engelbracht {et~al.}(2006)Engelbracht, Kundurthy, Gordon, Rieke,
  Kennicutt, Smith, Regan, Makovoz, Sosey, Draine, Helou, Armus, Calzetti,
  Meyer, Bendo, Walter, Hollenbach, Cannon, Murphy, Dale, Buckalew, \&
  Sheth}]{2006ApJ...642L.127E}
Engelbracht, C.~W., Kundurthy, P., Gordon, K.~D., {et~al.} 2006, The
  Astrophysical Journal, 642, L127

\bibitem[{Field {et~al.}(2011)Field, Blackman, \& Keto}]{2011MNRAS.416..710F}
Field, G.~B., Blackman, E.~G., \& Keto, E.~R. 2011, Monthly Notices of the
  Royal Astronomical Society, 416, 710

\bibitem[{Fielding {et~al.}(2020)Fielding, Ostriker, Bryan, \&
  Jermyn}]{2020ApJ...894L..24F}
Fielding, D.~B., Ostriker, E.~C., Bryan, G.~L., \& Jermyn, A.~S. 2020, The
  Astrophysical Journal Letters, 894, L24

\bibitem[{Gronke \& Oh(2018)}]{2018MNRAS.480L.111G}
Gronke, M., \& Oh, S.~P. 2018, arXiv.org, L111

\bibitem[{Harris {et~al.}(2020)Harris, Millman, Walt, Gommers, Virtanen,
  Cournapeau, Wieser, Taylor, Berg, Smith, Kern, Picus, Hoyer, Kerkwijk, Brett,
  Haldane, R{\'\i}o, Wiebe, Peterson, G{\'e}rard-Marchant, Sheppard, Reddy,
  Weckesser, Abbasi, Gohlke, \& Oliphant}]{Harris:2020fx}
Harris, C.~R., Millman, K.~J., Walt, S.~J., {et~al.} 2020, Nature, 1

\bibitem[{Heckman {et~al.}(1990)Heckman, Armus, \& Miley}]{1990ApJS...74..833H}
Heckman, T.~M., Armus, L., \& Miley, G.~K. 1990, Astrophysical Journal
  Supplement Series (ISSN 0067-0049), 74, 833

\bibitem[{Hoopes {et~al.}(2005)Hoopes, Heckman, Strickland, Seibert, Madore,
  Rich, Bianchi, Gil~de Paz, Burgarella, Thilker, Friedman, Barlow, Byun,
  Donas, Forster, Jelinsky, Lee, Malina, Martin, Milliard, Morrissey, Neff,
  Schiminovich, Siegmund, Small, Szalay, Welsh, \& Wyder}]{2005ApJ...619L..99H}
Hoopes, C.~G., Heckman, T.~M., Strickland, D.~K., {et~al.} 2005, The
  Astrophysical Journal, 619, L99

\bibitem[{Kaneda {et~al.}(2010)Kaneda, Ishihara, Suzuki, Ikeda, Onaka,
  Yamagishi, Ohyama, Wada, \& Yasuda}]{2010A&A...514A..14K}
Kaneda, H., Ishihara, D., Suzuki, T., {et~al.} 2010, arXiv.org, A14

\bibitem[{Keto \& {Myers, P. C.}(1986)}]{1986ApJ...304..466K}
Keto, E.~R., \& {Myers, P. C.} 1986, Astrophysical Journal, 304, 466

\bibitem[{Kim {et~al.}(2020)Kim, Ostriker, Fielding, Smith, Bryan, Somerville,
  Forbes, Genel, \& Hernquist}]{2020ApJ...903L..34K}
Kim, C.-G., Ostriker, E.~C., Fielding, D.~B., {et~al.} 2020, The Astrophysical
  Journal Letters, 903, L34

\bibitem[{Krieger {et~al.}(2019)Krieger, Bolatto, Walter, Leroy, Zschaechner,
  Meier, Ott, Weiss, Mills, Levy, Veilleux, \& Gorski}]{2019ApJ...881...43K}
Krieger, N., Bolatto, A.~D., Walter, F., {et~al.} 2019, The Astrophysical
  Journal, 881, 43

\bibitem[{Krieger {et~al.}(2020)Krieger, Bolatto, Koch, Leroy, Rosolowsky,
  Walter, Wei{\ss}, Eden, Levy, Meier, Mills, Moore, Ott, Su, \&
  Veilleux}]{2020ApJ...899..158K}
Krieger, N., Bolatto, A.~D., Koch, E.~W., {et~al.} 2020, The Astrophysical
  Journal, 899, 158

\bibitem[{Larson(1981)}]{Larson:1981jma}
Larson, R.~B. 1981, Monthly Notices of the Royal Astronomical Society, 194, 809

\bibitem[{Leeuw \& Robson(2009)}]{2009AJ....137..517L}
Leeuw, L.~L., \& Robson, E.~I. 2009, The Astronomical Journal, 137, 517

\bibitem[{Lehnert {et~al.}(1999)Lehnert, Heckman, \&
  Weaver}]{1999ApJ...523..575L}
Lehnert, M.~D., Heckman, T.~M., \& Weaver, K.~A. 1999, arXiv.org, 575

\bibitem[{Leroy {et~al.}(2015)Leroy, Walter, Martini, Roussel, Sandstrom, Ott,
  Wei{\ss}, Bolatto, Schuster, \& Dessauges-Zavadsky}]{2015ApJ...814...83L}
Leroy, A.~K., Walter, F., Martini, P., {et~al.} 2015, The Astrophysical
  Journal, 814, 83

\bibitem[{Li {et~al.}(2020)Li, Wang, Wu, Ma, \& Lin}]{2020RAA....20...31L}
Li, C., Wang, H.-c., Wu, Y.-w., Ma, Y.-h., \& Lin, L.-h. 2020, Research in
  Astronomy and Astrophysics, 20, 031

\bibitem[{Lopez {et~al.}(2020)Lopez, Mathur, Nguyen, Thompson, \&
  Olivier}]{2020ApJ...904..152L}
Lopez, L.~A., Mathur, S., Nguyen, D.~D., Thompson, T.~A., \& Olivier, G.~M.
  2020, The Astrophysical Journal, 904, 152

\bibitem[{Martini {et~al.}(2018)Martini, Leroy, Mangum, Bolatto, Keating,
  Sandstrom, \& Walter}]{2018ApJ...856...61M}
Martini, P., Leroy, A.~K., Mangum, J.~G., {et~al.} 2018, The Astrophysical
  Journal, 856, 61

\bibitem[{McKeith {et~al.}(1993)McKeith, Castles, Greve, \&
  Downes}]{1993A&A...272...98M}
McKeith, C.~D., Castles, J., Greve, A., \& Downes, D. 1993, Astronomy {\&}
  Astrophysics, 272, 98

\bibitem[{McKeith {et~al.}(1995)McKeith, Greve, Downes, \&
  Prada}]{1995A&A...293..703M}
McKeith, C.~D., Greve, A., Downes, D., \& Prada, F. 1995, Astronomy {\&}
  Astrophysics, 293, 703

\bibitem[{McMullin {et~al.}(2007)McMullin, Waters, Schiebel, Young, \&
  Golap}]{McMullin:2007tj}
McMullin, J.~P., Waters, B., Schiebel, D., Young, W., \& Golap, K. 2007,
  Astronomical Data Analysis Software and Systems XVI, 376, 127

\bibitem[{Nakai {et~al.}(1987)Nakai, Hayashi, Handa, Sofue, Hasegawa, \&
  Sasaki}]{1987PASJ...39..685N}
Nakai, N., Hayashi, M., Handa, T., {et~al.} 1987, Astronomical Society of
  Japan, 39, 685

\bibitem[{R~C~Kennicutt {et~al.}(2003)R~C~Kennicutt, Armus, Bendo, Calzetti,
  Dale, Draine, Engelbracht, Gordon, Grauer, Helou, Hollenbach, Jarrett,
  Kewley, Leitherer, Li, Malhotra, Regan, Rieke, Rieke, Roussel, Smith,
  Thornley, \& Walter}]{2003PASP..115..928K}
R~C~Kennicutt, J., Armus, L., Bendo, G., {et~al.} 2003, arXiv.org, 928

\bibitem[{Ranalli {et~al.}(2008)Ranalli, Comastri, Origlia, \&
  Maiolino}]{2008MNRAS.386.1464R}
Ranalli, P., Comastri, A., Origlia, L., \& Maiolino, R. 2008, Monthly Notices
  of the Royal Astronomical Society, 386, 1464

\bibitem[{Roussel {et~al.}(2010)Roussel, Wilson, Vigroux, Isaak, Sauvage,
  Madden, Auld, Baes, Barlow, Bendo, Bock, Boselli, Bradford, Buat,
  Castro-Rodriguez, Chanial, Charlot, Ciesla, Clements, Cooray, Cormier,
  Cortese, Davies, Dwek, Eales, Elbaz, Galametz, Galliano, Gear, Glenn, Gomez,
  Griffin, Hony, Levenson, Lu, O'Halloran, Okumura, Oliver, Page, Panuzzo,
  Papageorgiou, Parkin, Perez-Fournon, Pohlen, Rangwala, Rigby, Rykala, Sacchi,
  Schulz, Schirm, Smith, Spinoglio, Stevens, Srinivasan, Symeonidis, Trichas,
  Vaccari, Wozniak, Wright, \& Zeilinger}]{2010A&A...518L..66R}
Roussel, H., Wilson, C.~D., Vigroux, L., {et~al.} 2010, Astronomy {\&}
  Astrophysics, 518, L66

\bibitem[{Salak {et~al.}(2013)Salak, Nakai, Miyamoto, Yamauchi, \&
  Tsuru}]{2013PASJ...65...66S}
Salak, D., Nakai, N., Miyamoto, Y., Yamauchi, A., \& Tsuru, T.~G. 2013,
  Publications of the Astronomical Society of Japan, 65, 66

\bibitem[{Scannapieco \& Br{\"u}ggen(2015)}]{2015ApJ...805..158S}
Scannapieco, E., \& Br{\"u}ggen, M. 2015, The Astrophysical Journal, 805, 158

\bibitem[{Schneider \& Robertson(2017)}]{2017ApJ...834..144S}
Schneider, E.~E., \& Robertson, B.~E. 2017, The Astrophysical Journal, 834, 144

\bibitem[{Seaquist \& Clark(2001)}]{2001ApJ...552..133S}
Seaquist, E.~R., \& Clark, J. 2001, The Astrophysical Journal, 552, 133

\bibitem[{Shopbell \& Bland-Hawthorn(1998)}]{1998ApJ...493..129S}
Shopbell, P.~L., \& Bland-Hawthorn, J. 1998, The Astrophysical Journal, 493,
  129

\bibitem[{Sparre {et~al.}(2020)Sparre, Pfrommer, \&
  Ehlert}]{2020MNRAS.499.4261S}
Sparre, M., Pfrommer, C., \& Ehlert, K. 2020, Monthly Notices of the Royal
  Astronomical Society, 499, 4261

\bibitem[{Strickland \& Heckman(2007)}]{2007ApJ...658..258S}
Strickland, D.~K., \& Heckman, T.~M. 2007, The Astrophysical Journal, 658, 258

\bibitem[{Taylor {et~al.}(2001)Taylor, Walter, \& Yun}]{2001ApJ...562L..43T}
Taylor, C.~L., Walter, F., \& Yun, M.~S. 2001, The Astrophysical Journal, 562,
  L43

\bibitem[{Veilleux {et~al.}(2020)Veilleux, Maiolino, Bolatto, \&
  Aalto}]{2020A&ARv..28....2V}
Veilleux, S., Maiolino, R., Bolatto, A.~D., \& Aalto, S. 2020, The Astronomy
  and Astrophysics Review, 28, 2

\bibitem[{Veilleux {et~al.}(2009)Veilleux, Rupke, \&
  Swaters}]{2009ApJ...700L.149V}
Veilleux, S., Rupke, D. S.~N., \& Swaters, R. 2009, The Astrophysical Journal
  Letters, 700, L149

\bibitem[{Virtanen {et~al.}(2020)Virtanen, Gommers, Oliphant, Haberland, Reddy,
  Cournapeau, Burovski, Peterson, Weckesser, Bright, van~der Walt, Brett,
  Wilson, Millman, Mayorov, Nelson, Jones, Kern, Larson, Carey, Polat, Feng,
  Moore, VanderPlas, Laxalde, Perktold, Cimrman, Henriksen, Quintero, Harris,
  Archibald, Ribeiro, Pedregosa, van Mulbregt, \&
  Contributors}]{2020NatMe..17..261V}
Virtanen, P., Gommers, R., Oliphant, T.~E., {et~al.} 2020, Nature Methods, 17,
  261

\bibitem[{Walter {et~al.}(2002)Walter, Wei{\ss}, \&
  Scoville}]{2002ApJ...580L..21W}
Walter, F., Wei{\ss}, A., \& Scoville, N. 2002, The Astrophysical Journal, 580,
  L21

\bibitem[{Westmoquette {et~al.}(2009)Westmoquette, Smith, Gallagher, Trancho,
  Bastian, \& Konstantopoulos}]{2009ApJ...696..192W}
Westmoquette, M.~S., Smith, L.~J., Gallagher, J. S.~I., {et~al.} 2009, The
  Astrophysical Journal, 696, 192

\bibitem[{Yun {et~al.}(1993)Yun, Ho, \& Lo}]{1993ApJ...411L..17Y}
Yun, M.~S., Ho, P. T.~P., \& Lo, K.~Y. 1993, Astrophysical Journal, 411, L17

\end{thebibliography}
\bibliographystyle{aasjournal}

\end{document}